\documentclass[conference]{IEEEtran}

\ifCLASSINFOpdf
\else
\fi

\usepackage{fancyhdr}
\usepackage{algorithm}
\usepackage{xspace}
\usepackage[font=small,labelfont=bf]{caption}
\usepackage{amsmath}
\usepackage{mathtools}
\usepackage{xcolor}
\usepackage{amsfonts}
\usepackage{amssymb}
\usepackage{pifont}
\usepackage{color}
\usepackage[normalem]{ulem}
\usepackage[hyphens]{url}
\usepackage[sort,nocompress]{cite}
\usepackage[final]{microtype}
\usepackage{flushend}
\usepackage{makecell}
\usepackage[bookmarks=true,breaklinks=true,letterpaper=true,colorlinks,linkcolor=black,citecolor=blue,urlcolor=black]{hyperref}
\usepackage{graphicx}
\usepackage{textcomp}
\usepackage{multirow}
\usepackage{textgreek}
\usepackage{subfig}
\usepackage{soul}
\usepackage{colortbl}
\usepackage{tabulary}
\usepackage{etoolbox}
\usepackage{algpseudocode}
\usepackage[]{siunitx}

\makeatletter
\algrenewcommand\ALG@beginalgorithmic{\footnotesize}
\makeatother

\xspaceaddexceptions{]\}}

\newcommand{\old}[1]{}

\newcommand{\fig}[1]{Figure~\ref{#1}}
\newcommand{\sect}[1]{Section~\ref{#1}}
\newcommand{\tab}[1]{Table~\ref{#1}}

\newcommand{\pimsim}[0]{uPIMulator\xspace}
\newcommand{\upmempim}[0]{UPMEM-PIM\xspace}

\renewcommand{\baselinestretch}{.999}
\title{Pathfinding Future PIM Architectures by \\Demystifying a Commercial PIM Technology} 

\begin{document}

\author{
\IEEEauthorblockN{Bongjoon Hyun\hspace{1em} Taehun Kim\hspace{1em} Dongjae Lee\hspace{1em} Minsoo Rhu\hspace{1em}}


\IEEEauthorblockA{KAIST
\\\texttt{\{bongjoon.hyun, taehun.kim, dongjae.lee, mrhu\}@kaist.ac.kr}}
}
\maketitle

\pagestyle{plain}

\IEEEpeerreviewmaketitle

\begin{abstract}

Processing-in-memory (PIM) has been explored for decades by computer
architects, yet it has never seen the light of day in real-world products due
to its high design overheads and lack of a killer application.  With the advent
of critical memory-intensive workloads, several commercial PIM technologies
have been introduced to the market, ranging from domain-specific PIM
architectures to more general-purpose PIM architectures.  In this work, we
deepdive into UPMEM's commercial PIM technology, a general-purpose PIM-enabled
parallel computing architecture that is highly programmable.  Our first key
contribution is the development of a flexible simulation framework for PIM. The
simulator we developed (aka \pimsim) enables the compilation of \upmempim
source codes into its compiled machine-level instructions, which are
subsequently consumed by our cycle-level performance simulator.  Using \pimsim,
we demystify UPMEM's PIM design through a detailed characterization study.
Finally, we identify some key limitations of the current UPMEM-PIM system
through our case studies and present some important architectural features that
will become critical for future PIM architectures to support.

\end{abstract}

\section{Introduction}

\begin{quotation}
\noindent \em ``We’ve investigated applying PIM to our workloads and determined
there are several challenges to using these approaches. Perhaps the biggest
challenge of PIM is its programmability. It is hard to anticipate future model
compression methods, so programmability is required to adapt to these. PIM must
also support flexible parallelization since it is hard to predict how much each
dimension (of embedding tables) will scale in the future.'' 
  
  \emph{``First-Generation Inference Accelerator Deployment at Facebook''},
Facebook,
2021~\cite{first_generation_infernece_accelerator_deployment_at_facebook}
\end{quotation}
\vspace{1em}

Emerging workloads in the areas of scientific computing, graph processing, and
machine learning pose unprecedented demand for larger data. However, the
well-known memory \emph{bandwidth} wall causes a critical performance
bottleneck for these memory-bound workloads, due to the widening performance
gap between processor and memory.  Processing-in-memory (PIM) architectures
have been explored extensively for
decades~\cite{computational_ram,iram,active_pages,diva}, as they help alleviate
the memory bandwidth bottleneck by moving compute logic closer to memory.
Unfortunately, the computing industry has so far been lukewarm in
commercializing  PIM architectures, primarily because of their high design
overheads (e.g., regression in DRAM density, thermal
issues~\cite{namsungkim_thermal}) and their intrusiveness to the software stack
(e.g., programming model~\cite{a_modern_primer_on_processing_in_memory, tom,
scheduling_techniques_for_gpu_architectures_with_processing_in_memory_capabilities,
to_pim_or_not}, managing address space and data
coherence~\cite{pim_enabled_instructions, conda, lazypim}), rendering PIM
mostly an academic pursuit.

Interestingly, such sentiment towards PIM has changed dramatically over the
past couple of years with several commercial PIM systems introduced to the
market. These PIM designs can broadly be grouped into two categories: 1)
domain-specific PIM and 2) general-purpose PIM.  Domain-specific PIM designs
have been driven by key memory vendors like Samsung~\cite{hbm_pim_isca,
hbm_pim_isscc, axdimm,
aquabolt_xl_hbm2_pim_lpddr5_pim_with_in_memory_processing_and_axdimm_with_acceleration_buffer}
and SK Hynix~\cite{newton, aim}, which focus on specializing their PIM design
by supporting key compute primitives for a targeted application domain (e.g.,
matrix multiplication for accelerating deep neural networks), reigniting
people's interest in PIM designs~\cite{graph_p, graph_h, graph_q, scalagraph,
tesseract, hbm_pim_isca,
aquabolt_xl_hbm2_pim_lpddr5_pim_with_in_memory_processing_and_axdimm_with_acceleration_buffer,
newton, tetris, neurocube, tensor_dimm, recnmp, trim_micro, fafnir, impica,
polynesia, dracc, isaac, dimmining, spacea, gearbox, natsa, drisa, mcdram,
mcdram_v2, medal,tensor_casting,smartsage}.  At the other end of the spectrum,
the PIM solution offered by UPMEM~\cite{upmem} (henceforth referred to as
UPMEM-PIM) takes a different approach by providing a \emph{general-purpose}
parallel programming language with an LLVM-based compiler
stack~\cite{llvm,upmem_llvm}, providing application developers the flexibility
to write any parallel program to be executed using PIM.  Thanks to its high
programmability and flexibility, several recent work studied the
applicability of UPMEM-PIM for accelerating a variety of application domains,
e.g., graph algorithms, machine learning, bioinformatics, etc.~\cite{prim,
a_case_study_of_processing_in_memory_in_off_the_shelf_systems,
upmem_sigmod,upmem_rna_seq,bulk_jpeg_decoding_on_in_memory_processors}.
Similar to how GPUs have transformed themselves into a first-class computing
citizen after years of hardware/software refinement, we believe that it is
possible for such general-purpose PIM design to similarly evolve into an
important computing device (or at a minimum provide valuable insights in
designing future general-purpose PIM) as its hardware/software stack matures.

Given this landscape, our key motivation is to demystify industry's first
general-purpose PIM design through a detailed characterization study,
understanding the unique properties of  UPMEM-PIM and identifying important
research domains that computer architects can explore. To this end, we first
develop an \upmempim ISA compatible {simulation framework that utilizes UPMEM's
open-source compiler stack to compile \emph{any} \upmempim program, from its
C-level source code down to its machine level instructions. The compiled
\upmempim binary is then consumed by our cycle-level hardware performance
simulator, which we carefully cross-validate against a real \upmempim system
(\sect{sect:sim_framework}).  Simulators are, by design, immensely flexible and
customizable, so they enable us to understand the fine-grained details of the
runtime execution of a \emph{real} (UPMEM) PIM program. Using our PIM simulator
(henceforth referred to as UPMEM-PIM simulator, aka \pimsim), we conduct a
workload characterization study and provide a number of interesting insights
that cannot be easily uncovered using \upmempim chip’s hardware performance
counters or profiling tools (\sect{sect:workload_characterization}).  Finally,
we uncover some critical limitations of the current UPMEM-PIM system through
our case studies and propose several key architectural features required for
PIM to become more performant, robust, and secure (\sect{sect:case_studies}).
These features include the need for vector processing and ILP
(instruction-level parallelism) enhancing microarchitectures, architectural
support for multi-tenant execution, and the support for on-demand caching
rather than solely relying on scratchpads.  Overall, we expect our in-depth
exploration of UPMEM-PIM using our \pimsim to open up important research
directions for computer system designers\footnote{\pimsim is open-sourced
at
\href{https://github.com/VIA-Research/uPIMulator}{https://github.com/VIA-Research/uPIMulator}.},
paving the way for PIM to evolve into a truly general-purpose computing device. 
		
\section{UPMEM-PIM Architecture} \label{sect:background}

\subsection{Hardware Architecture}
\label{sect:background_hw_arch}

\begin{figure}[t!] \centering
  \includegraphics[width=0.485\textwidth]{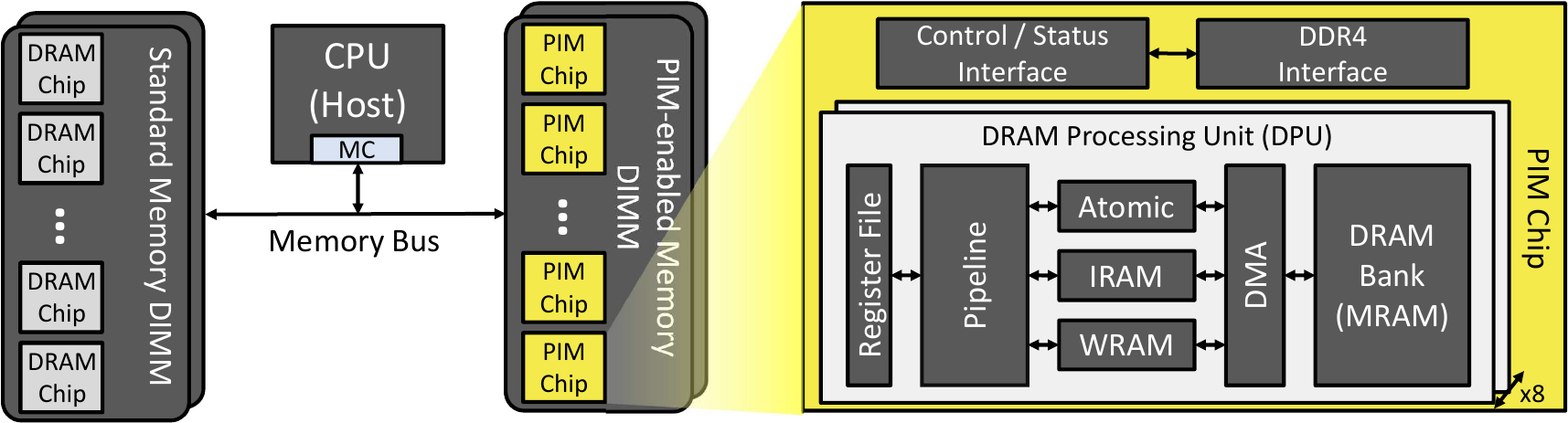} 
  \caption{\upmempim hardware system overview.}
  \label{fig:upmem_pim_enabled_memory_overview}
  \vspace{-1em}
\end{figure}

{\bf System overview.} \fig{fig:upmem_pim_enabled_memory_overview} provides a
high-level overview of an \upmempim based system containing a host-side CPU
communicating with a group of standard regular DIMMs and another group of
PIM-enabled memory DIMMs (\upmempim modules). An \upmempim module is based on a
standard DDR4-2400~\cite{ddr4_2400} DIMM form factor containing $8$ \upmempim
DRAM chips per each rank. Within each \upmempim DRAM chip, there are $8$ DPUs
(DRAM Processing Units), one DPU per each DRAM bank. Each DPU has direct access
to a dedicated $64$ MB DRAM bank (referred to as \emph{Main RAM}, aka MRAM), a
$64$ KB SRAM-based scratchpad memory (aka \emph{Working RAM}, WRAM), and $24$
KB instruction memory (aka \emph{Instruction RAM}, IRAM).  Before an \upmempim
program (i.e., the PIM \emph{kernel}) is to be executed, the host CPU must
explicitly offload both the PIM kernel and the input data from CPU's
conventional memory address space (i.e., regular DIMMs) to DPU's \upmempim
address space.  The real PIM system we study in this work contains $20$
double-ranked \upmempim DIMMs, so a total of
($20\times2\times8\times8$)$=$2,560 DPUs concurrently execute as co-processors
to the CPU.

\begin{figure}[t!] \centering
\subfloat[Host-side code.]{\includegraphics[width=0.495\textwidth]{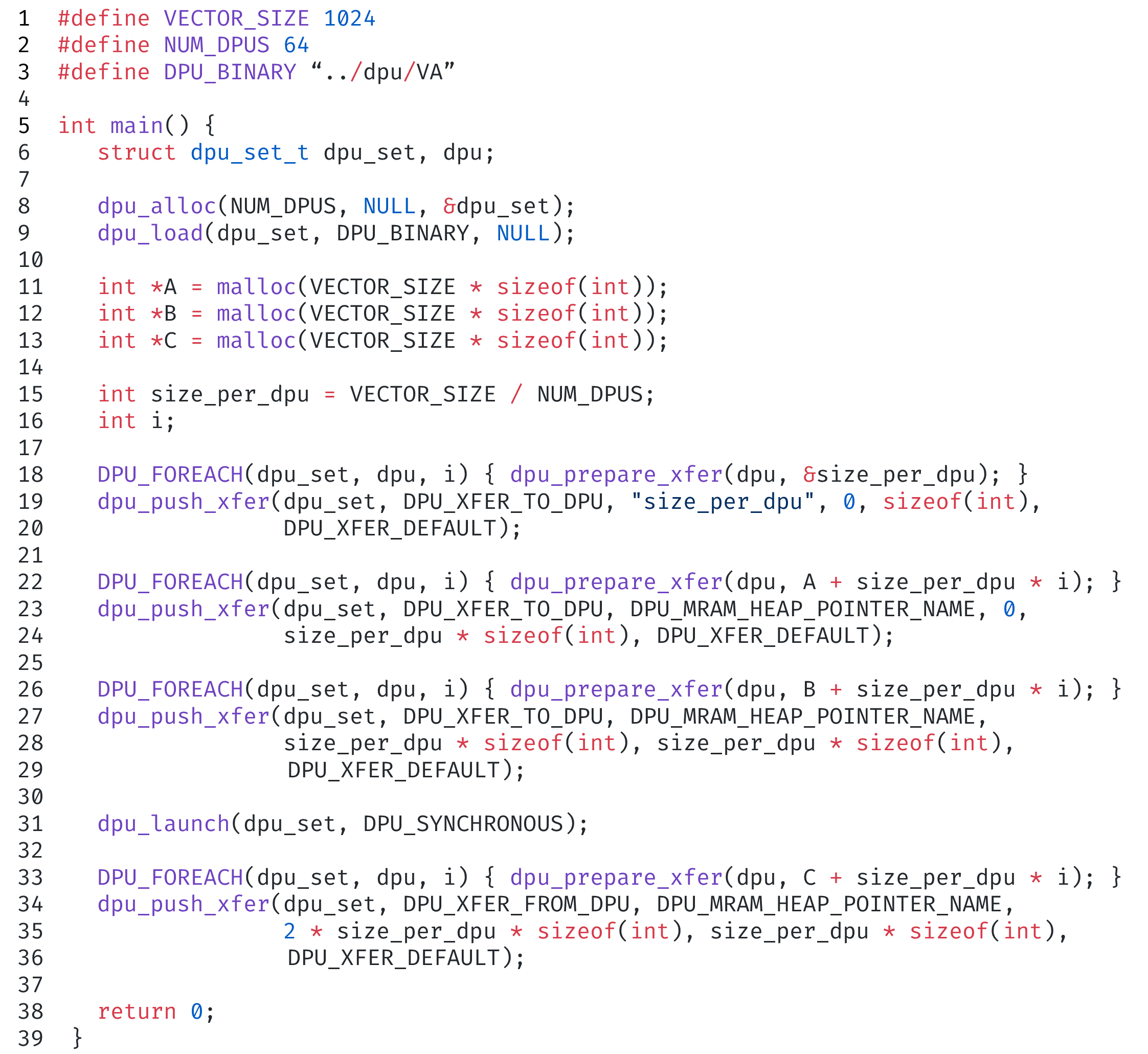}\label{fig:upmem_host}}\\
\subfloat[DPU-side code.]{\includegraphics[width=0.495\textwidth]{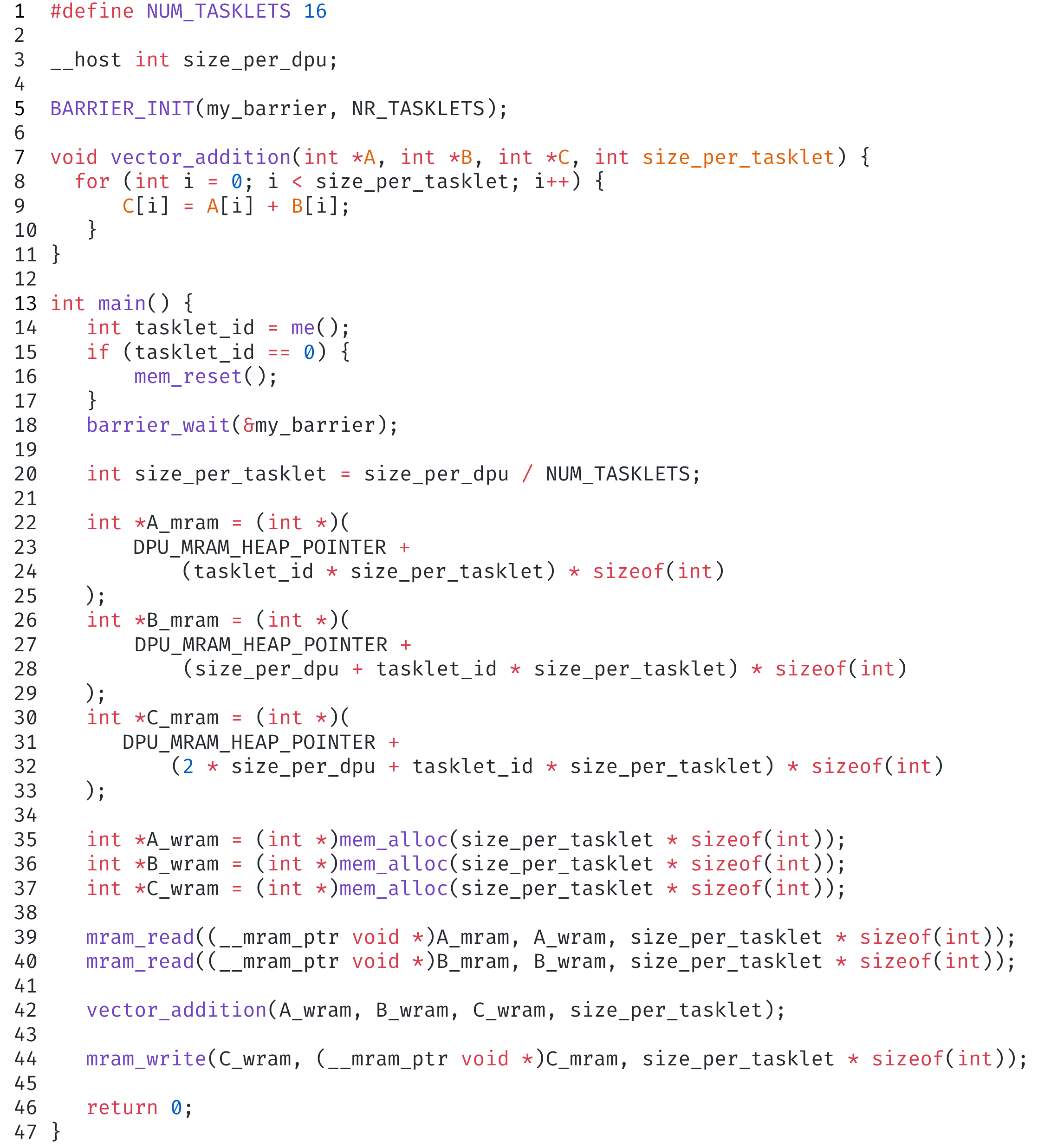}\label{fig:upmem_dpu}}
\caption{An element-wise vector addition program written for \upmempim: (a) host-side and (b) DPU-side program.}
\label{fig:upmem_programming}
\vspace{-1em}
\end{figure}

{\bf DPU architecture.} The DPU is designed as an in-order 14-stage pipelined
processor with a RISC-based ISA, implementing fine-grained multi-threading. A
total of $24$  threads (called \emph{tasklets} by UPMEM) can concurrently
execute within a DPU, all of which share the scratchpad (WRAM), instruction
memory (IRAM), and per-bank DRAM (MRAM).  The UPMEM DPU has a peculiar thread
scheduling rule where two consecutive instructions within the same thread must
be dispatched $11$ cycles apart (UPMEM refers to such microarchitecture as the
\emph{revolver pipeline}~\cite{upmem_isa}).  UPMEM states that such scheduling
constraint is enforced to obviate the need to implement complicated circuitry
for data forwarding and pipeline interlocks~\cite{upmem_hotchips}.   Another
unique aspect of the DPU microarchitecture is in its register file (RF) design:
the RF is split into an even and odd RF and a thread cannot access multiple
even or odd registers at the same cycle (e.g., r0 and r2 cannot be accessed at
the same cycle) due to a structural hazard (i.e., RF conflict). 

\subsection{Programming Model}
\label{sect:background_pl}

\upmempim follows the single-program multiple-data (SPMD) programming paradigm.
A single program written by the programmer gets executed by all the software
threads (i.e., tasklets) that are instantiated, but each individual thread can
take its own control flow and access different parts of the data using its
thread ID (tasklet ID).  Since there can be up to 2,560 DPUs and $24$ threads
per DPU, the programmer must carefully partition the input data, not only
across the DPUs (\fig{fig:upmem_programming}(a), line $18$-$20$, $22$-$24$, and
$26$-$29$) but also across the threads within each DPU
(\fig{fig:upmem_programming}(b), line $22$-$29$).  We use
\fig{fig:upmem_programming} as a running example to highlight some of the
important programming semantics of \upmempim.

{\bf Host-side programming.} Any program that is written in UPMEM's C-like
programming language can be compiled into its machine code by using the
LLVM-based compiler toolchain~\cite{llvm} developed by UPMEM~\cite{upmem_llvm}.
Similar to NVIDIA's CUDA~\cite{cuda}, \upmempim follows the \emph{co-processor}
computing model where the CPU \emph{offloads} the memory-intensive task to the
DPU, functioning as an arbiter of the PIM program's execution. Consequently,
the UPMEM compiler generates two binaries, one that runs on the host and the
other that runs across all the DPUs. In the host-side code
(\fig{fig:upmem_programming}(a)), the programmer must ($1$) allocate the
desired number of DPUs to be used (line $8$), ($2$) offload the program binary
to all the DPUs (line $9$), ($3$) partition and send input data to the DPU's
scratchpad (line $18$-$20$) and per-bank DRAM (line $22$-$24$, and $26$-$29$),
($4$) ask the host to send commands to the DPUs to execute the PIM program
(line $31$), and ($5$) once the PIM program terminates, retrieve back the
results from DPU memory back to the host CPU's memory address space (line
$33$-$36$). 

{\bf DPU-side programming.} A unique aspect of \upmempim's programming model is
that all the PIM kernel's working set \emph{must} be staged through DPU's
scratchpad using DMA instructions. Consider the code snippet in
\fig{fig:upmem_programming}(b). Any thread executing within the DPU can only
load (store) data from (to) the scratchpad (WRAM) address space  but it is not
able to address data in the per-bank DRAM (MRAM) address space directly (line
$7$, $42$). 	In effect, DPUs operate over \emph{two} distinct memory address
spaces, the slower but larger per-bank DRAM region and the faster yet smaller
scratchpad region.  Only when the programmer explicitly requests data movements
from the per-bank DRAM region to the scratchpad region (using DMA instructions
via \texttt{mram\_read()}, line $39$-$40$) can the DPU threads access the
necessary data from the scratchpad using load/store instructions (line $9$,
notice the pointers to the arrays $A$,$B$,$C$ are dynamically allocated at the
scratchpad WRAM via \texttt{mem\_alloc()} calls in line $35$-$37$). This is
similar to NVIDIA's CUDA programming model~\cite{cuda} where the programmer
must explicitly orchestrate data movements across the CPU memory and the GPU
memory using \texttt{cudaMemcpy()} (unless the programmer employs Unified
Memory~\cite{cuda_uvm}). CUDA, however, does allow threads to directly load
(store) from (to) \emph{both} its scratchpad and its DRAM, unlike UPMEM's
memory model which only allows load/store semantics over the scratchpad
(\fig{fig:memory_model_cuda_vs_upmem}(a,b)).  In the remainder of this paper,
we refer to such a model as UPMEM's \emph{scratchpad-centric} programming
model.

\begin{figure}[t!] \centering
  \includegraphics[width=0.485\textwidth]{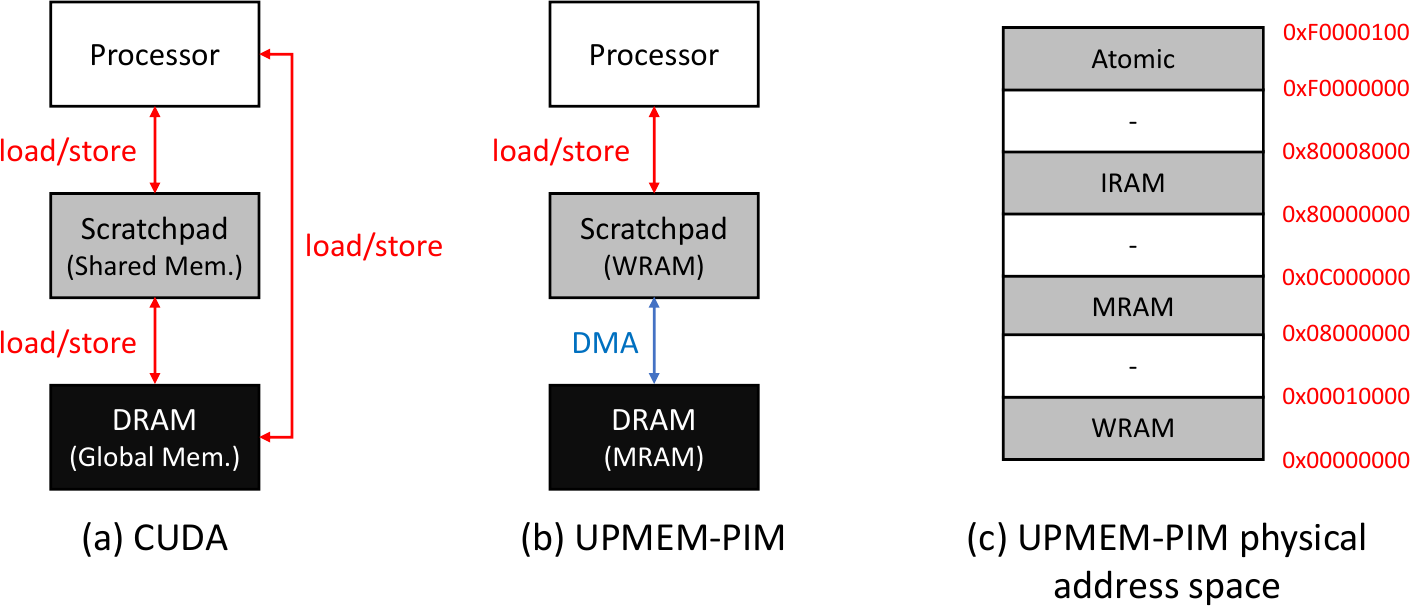}
  \caption{Memory model of (a) CUDA and (b) \upmempim. (c) The (physical) address map of \upmempim. }
  \label{fig:memory_model_cuda_vs_upmem}
  \vspace{-1em}
\end{figure}

\begin{figure*}[t!] \centering
  \includegraphics[width=0.99\textwidth]{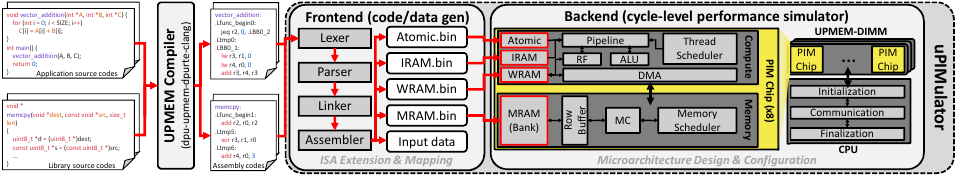}
\caption{\pimsim simulation framework overview.}
\vspace{-1em}
\label{fig:simulation_framework}
\end{figure*}

{\bf Data sharing and synchronization.} Threads executing within the same DPU
can share data over the DPU scratchpad or its local DRAM bank (MRAM). They can
also synchronize with each other by using mutexes, barriers, or semaphores
allocated in \upmempim's atomic memory region
(\fig{fig:memory_model_cuda_vs_upmem}(c)), all of which are supported by
UPMEM's SDK~\cite{upmem_sdk}. 

However, threads executing in different DPUs cannot share data or synchronize
with each other directly. This is because 1) all the DPUs' per-bank DRAM across
the \upmempim DIMM are not virtualized within a single, \emph{shared} memory
address space (further discussed in \sect{sect:system_sw}) 2) nor is there a
direct communication datapath among them.  If data sharing or synchronization
across different DPUs is in need, the programmer must first explicitly copy
back the shared data from the producer DPU's memory to the CPU memory after
kernel terminates. The CPU can then copy back this shared data from its CPU
memory region to the consumer DPU's memory region during the next PIM kernel
execution.

\subsection{System Software for Memory Management}
\label{sect:system_sw}

\upmempim does not have a memory management unit (MMU) to virtualize its
physical memory, so the DPU uses \emph{physical} addresses when accessing WRAM,
IRAM, and MRAM, as illustrated in \fig{fig:memory_model_cuda_vs_upmem}(c).  In
other words, when moving data across \upmempim's memory hierarchy using (1)
load/store instructions (for scratchpad$\leftrightarrow$RF) or (2) DMA
instructions (for DRAM$\leftrightarrow$scratchpad), the memory addresses
generated by executing an instruction are used \emph{as-is}, without any
address translation process involved (\fig{fig:memory_model_cuda_vs_upmem}(b)).
Consider the example in \fig{fig:upmem_programming}. When the input array $B$
is being copied from the CPU to DPU's per-bank DRAM
(\fig{fig:upmem_programming}(a), line $26$-$29$) and then from DPU's DRAM to
DPU's scratchpad (\fig{fig:upmem_programming}(b), line $40$), the programmer
must carefully program the pointer value to use as the destination (for
CPU$\rightarrow$DPU's DRAM) and source address (for DPU's
DRAM$\rightarrow$DPU's scratchpad) within per-bank DRAM (MRAM) by using
\texttt{DPU\_MRAM\_HEAP\_POINTER\_NAME} (\fig{fig:upmem_programming}(a), line
$27$-$29$) or \texttt{DPU\_MRAM\_HEAP\_POINTER}
(\fig{fig:upmem_programming}(b), line $26$-$29$) as the base physical address.

Overall, the lack of a virtual memory support leaves the programmer with the
burden of reasoning about where the input (output) data should be copied over
to (from) within DPU's DRAM, hurting user productivity.
\sect{sect:case_multitenancy} further discusses the architectural implication
of an MMU-less PIM.

\section{\pimsim Simulation Framework} \label{sect:sim_framework}

\fig{fig:simulation_framework}  provides an overview of \pimsim, which consists
of two key components: (1) a compiler toolchain that supports execution-driven
simulation of UPMEM ISA-compatible, machine-level instructions, and (2) a
hardware performance simulator cross-validated against a real \upmempim.
Together, these dual components reduce the effort required to model UPMEM's
general-purpose PIM architecture with high accuracy, enabling architectural
exploration of any PIM program written with UPMEM's programming model. 

\subsection{Simulator Development}
\label{sect:sim_dev}

{\bf Software compilation toolchain.} The open-source UPMEM
SDK~\cite{upmem_sdk} provides an LLVM~\cite{llvm}-based compiler
toolchain~\cite{upmem_llvm} (\texttt{dpu-upmem-dpurte-clang}) that takes in (1)
the programmer-written  source codes and (2) \texttt{glibc}-style, \upmempim
compatible C library (e.g., \texttt{mem\_alloc()} for \texttt{malloc} in DPU
scratchpad, \texttt{memcpy()}, \texttt{printf()}) to preprocess, compile, and
assemble into  binary objects, finally linking them into an \upmempim binary
executable. \pimsim utilizes UPMEM SDK's preprocessor and compiler \emph{as-is}
to first lower UPMEM program source files into multiple assembly-level codes.
These assembly codes are then fed into our custom-designed linker (which is
based on the open-source ANTLR's lexer and
parser~\cite{the_definitive_antlr_4_reference, antlr, antlr_repo}) to go
through the lexical and syntax analyses to resolve  the def-use relationships
of all the functions, code labels, etc. for linking.  Finally, our
custom-designed assembler generates the final binary files to upload into
\upmempim's atomic (i.e., mutex), IRAM (i.e., the \upmempim program), WRAM, and
MRAM (i.e., the input data) address spaces (\fig{fig:simulation_framework}). 

The reason why \pimsim employs a custom-designed linker and assembler is as
follows. We observe that the current version of UPMEM linker is specifically
tied to \upmempim's microarchitecture, preventing us from exploring alternative
PIM architectures.  For instance, UPMEM's linker generates a linking error when
the compiled program's instruction memory or scratchpad usage exceeds the
physical IRAM or WRAM capacity.  As detailed later in \sect{sect:case_caching},
this paper presents a case study to evaluate the trade-offs of employing an
on-demand cache for \upmempim, as opposed to UPMEM's current scratchpad-centric
design.  Under UPMEM's programming model, this requires us to write the
\upmempim program that has a working set allocated in the scratchpad (WRAM)
space  exceeding its $64$ KB size, which is subsequently re-mapped to the
per-bank DRAM region in our cycle-level hardware performance simulator.  This
allows us to treat a DPU thread's load/store transactions to scratchpad as if
they are to DRAM, so plugging in a cache simulator in between the DPU and
scratchpad (which is emulated as DRAM) enables us to study the performance of
caches vs. scratchpads (\sect{sect:case_caching} details our methodology for
this study).  None of these features are available with UPMEM's current linker
design, motivating us to implement our own linker and assembler for a flexible
simulator development and design space exploration.

Overall, \pimsim enjoys LLVM's mature compiler stage optimizations (e.g.,
common subexpression elimination~\cite{compilers}) by leveraging UPMEM's
existing preprocessor/compiler as-is while also enabling diverse architectural
explorations through our custom-designed linker/assembler.

\sethlcolor{green}

\begin{table}[t!] \centering
\vspace{1em}
  \footnotesize
    \caption{ \pimsim default configuration. }
    \begin{tabular}{|c|c|}
      \hline
      \multicolumn{2}{|c|}{\textbf{DPU processor architecture}} \\
      \hline
      Operating frequency & $350$ MHz \\ \hline              
      Number of pipeline stages & $14$ \\ \hline              
      Revolver scheduling cycles & $11$ \\ \hline
      WRAM / IRAM size & $64$ KB / $24$ KB  \\ \hline

			WRAM / IRAM access latency & $1$ cycle \\ \hline
			WRAM / IRAM access granularity & $4$ / $6$ B per clock \\ \hline
			WRAM / IRAM access bandwidth & 1,400 / 2,100 MB/sec \\ \hline
          Atomic memory size & $256$ Bits \\ \hline
      \multicolumn{2}{|c|}{\textbf{DRAM system}} \\
      \hline
      MRAM  size &  $64$ MB \\ \hline
      DDR specification & DDR4-2400~\cite{ddr4_2400} \\ \hline
      Memory scheduling policy & FR-FCFS \\ \hline
      Row buffer size & $1$ KB \\ \hline
      tRCD, tRAS, tRP, tCL, tBL & $16$, $39$, $16$, $16$, $4$ cycles \\ \hline
      \multicolumn{2}{|c|}{\textbf{Communication}} \\
      \hline
      CPU$\rightarrow$DPU bandwidth (per rank) & 0.296 GB/s per DPU \\ \hline
      CPU$\leftarrow$DPU bandwidth (per rank) & 0.063 GB/s per DPU \\ \hline
      \multicolumn{2}{|c|}{\textbf{Software architecture}} \\
      \hline
      Number of general-purpose registers & 24 \\ \hline
      Maximum number of threads & 24 \\ \hline
      Stack size (per thread) & 2 KB \\ \hline
      Heap size & 4 KB \\ \hline
			\end{tabular}
\vspace{-1.3em}
  \label{tab:simulator_configuration}
  \end{table}

{\bf Hardware performance simulator.} We implement a cycle-level performance
simulator of  UPMEM DPU by referring to both UPMEM's user manual and publicly
available information and discussion about the DPU's
microarchitecture~\cite{upmem_hotchips, upmem_sdk, upmem_llvm, upmem_isa,
upmem_dpu_abi, upmem_white_paper}. As summarized in
\tab{tab:simulator_configuration}, the baseline DPU architecture is modeled as
a 14-stage in-order pipelined processor, faithfully modeling its revolver
pipeline scheduling algorithm and the structural hazard enforced at the
odd/even RF accesses (\sect{sect:background_hw_arch}). \pimsim functionally
executes the instructions to update its architectural state, allowing us to
verify the correctness of PIM program's execution. 

As for modeling the DRAM subsystem, rather than employing a highly accurate
cycle-level DRAM simulator~\cite{dramsim2,ramulator,usimm}, we base our
implementation on GPGPU-Sim's cycle-level DRAM simulator for fast simulation
time~\cite{gpgpu_sim} \sethlcolor{yellow} (our simulator runs $2.5\times$
slower when interfaced with Ramulator~\cite{ramulator}, which is known to be
the fastest among popular DRAM simulators).  Because the details of \upmempim's
memory scheduling policy is not publicly available, we employ a first-row,
first-come-first-serve (FR-FCFS~\cite{fr_fcfs})  algorithm for scheduling
memory transactions.  The communication latency of transferring data over the
CPU$\leftrightarrow$DPU channel is simulated by employing a fixed bandwidth
model as its communication channel (i.e., communication latency $=$ transfer
size/communication bandwidth), whose value is carefully tuned by profiling a
real \upmempim system (\tab{tab:simulator_configuration}).

Because \upmempim implements the CPU$\leftrightarrow$DPU communication using
Intel AVX read (CPU$\leftarrow$DPU) and write (CPU$\rightarrow$DPU)
instructions~\cite{intel_avx}, we observe asymmetric CPU$\leftrightarrow$DPU
communication bandwidth (i.e., the synchronous AVX reads have lower throughput
than the asynchronous AVX writes), a phenomenon also reported in \cite{prim}.

\subsection{Simulator Availability and Extensibility}
\label{sect:sim_extensibility}

\pimsim is designed to cleanly decouple the SPMD-based frontend code/data
generation from the backend performance model with its modular design
(\fig{fig:simulation_framework}).  Such design philosophy is inspired by
GPGPU-Sim~\cite{gpgpu_sim} which similarly utilizes NVIDIA's CUDA compiler and
PTX assembler as its frontend to generate CUDA code/data, which is subsequently
consumed by its backend cycle-level GPU microarchitecture simulator.  As such,
\pimsim can easily be extended to model and evaluate architecture designs with
alternative software/hardware architectures (we later demonstrate \pimsim's
extensibility via our case study in \sect{sect:case_studies}).  For instance,
one can modify \pimsim's frontend code/data generation stage to flexibly map
the code/data binaries at arbitrary locations in the memory address space, a
feature we utilize to generate the proper instructions/data in our ``cache vs.
scratchpad'' case study discussed later in \sect{sect:case_caching}.
Similarly, \pimsim's backend performance model can also be extended to execute
\upmempim's SPMD code over alternative hardware architectures. For instance,
one can maintain the same \upmempim's code to execute over an NVIDIA GPU style
SIMD processor architecture by modifying the backend performance model to be
implemented as a SIMT (single-instruction-multiple-thread)~\cite{cuda} vector
processor microarchitecture model, a case study we conduct in
\sect{sect:case_simt}.

\begin{table}[t!] \centering
\footnotesize
\caption{ PrIM benchmarks configurations used for the characterization and case studies conducted in this work.}
\begin{tabular}{|c|c|c|}
\hline
\textbf{Benchmark}  & \textbf{Dataset (single DPU)}     & \textbf{Dataset (multiple DPUs)}  \\ \hline
BFS                 & 2K vertices, 15K edges    & 16K vertices, 120K edges  \\ \hline
BS                  & 32K elem., 4K queries    & 128K elem., 16K queries  \\ \hline
GEMV                & 2K x 64, 64 x 1 elem.   & 8K x 64, 64 x 1 elem.  \\ \hline
HST-L               & 128K elem., 256 bins     & 512K elem., 256 bins     \\ \hline
HST-S               & 128K elem., 256 bins     & 512K elem., 256 bins     \\ \hline
MLP                 & 3 layers, 256 neurons    & 3 layers, 1K neurons    \\ \hline
NW           & 256 gene sequence   & 512 gene sequence \\ \hline
RED                 & 512K elem.                 & 2M elem.                 \\ \hline
SCAN-RSS            & 256K elem.                  & 1M elem.                  \\ \hline
SCAN-SSA            & 256K elem.                  & 1M elem.                  \\ \hline
SEL                 & 512K elem.                  & 2M elem.                 \\ \hline
SpMV                & 12K x 12K, 80519 elem.   & 14K x 14K, 316740 elem.  \\ \hline
TRNS                & 128K elem.                  & 256K elem.                \\ \hline
TS                  & 2K elem., 64 queries       & 64K elem., 64 queries  \\ \hline
UNI                 & 512K elem.                 & 2M elem.                 \\ \hline
VA                  & 1M elem.                  & 4M elem.                 \\ \hline
\end{tabular}
\vspace{-1.5em}
\label{tab:prim_configurations}
\end{table}

\begin{figure*}[t!] \centering
  \includegraphics[width=0.99\textwidth]{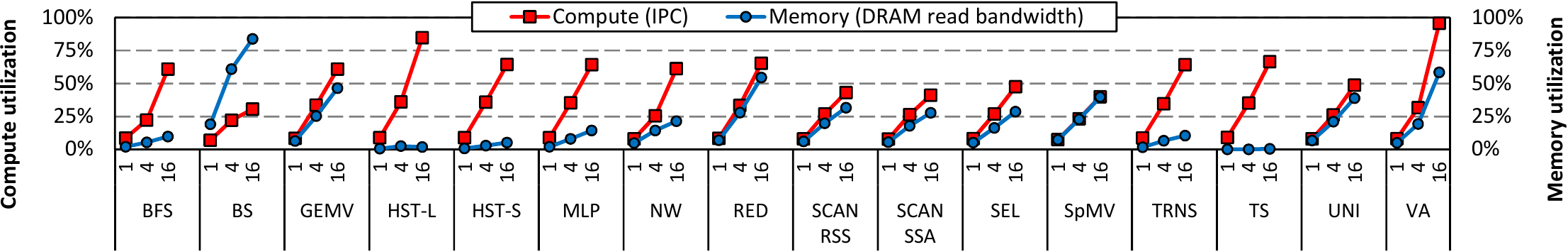}
\caption{PrIM's compute utilization (left axis) and memory read bandwidth utilization (right axis) when executing with $1$/$4$/$16$ threads. While a DPU's theoretical maximum DRAM bandwidth is 700 MB/sec, prior work~\cite{prim} observed that the maximum bandwidth is around $600$ MB/sec in real \upmempim system. We therefore configured \pimsim's DRAM bandwidth accordingly. 
A single DPU's max compute throughput is set as $1$ IPC and compute utilization is the percentage of this max IPC achieved.}
\vspace{-0.5em}
\label{fig:ipc_and_bw_utilization}
\end{figure*}

\begin{figure*}[t!] \centering
  \includegraphics[width=0.99\textwidth]{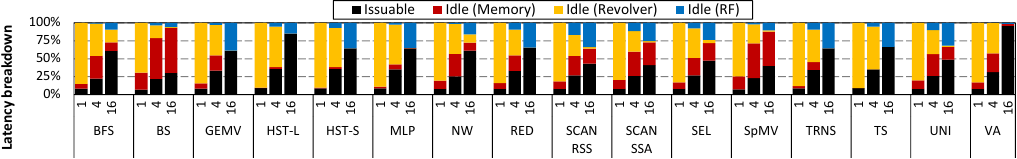}
\caption{Breakdown of DPU's runtime into active (black) and idle (red, yellow, blue) cycles. When all the threads are idle, we  categorize each thread's status based on the reason for its idleness, i.e., memory (red), revolver pipeline scheduling constraint (yellow), and the structural hazard at the RF (blue).}
\vspace{-1em}
\label{fig:single_dpu_latency_breakdown}
\end{figure*}

\subsection{Simulator Validation}
\label{sect:sim_validation}

We validate our \pimsim using PrIM~\cite{prim_repo}, an open-source \upmempim
benchmark suite (\tab{tab:prim_configurations}). PrIM consists of $16$
data-intensive workloads from various application domains such as  linear
algebra, graph processing, neural networks, etc. We verify \pimsim's functional
correctness as well as its performance correlation to real \upmempim hardware
by running each individual PrIM benchmark with $1$/$2$/$4$/$8$/$16$/$24$
threads under various input data sizes, cross-validating both \pimsim and real
\upmempim's final output data as well as its execution time.  Among the $16$
PrIM benchmarks, \pimsim was able to compile and simulate $13$ workloads as-is.
However, the remaining $3$ workloads (BFS, SpMV, NW) had minor bugs or utilized
undisclosed functions within the UPMEM SDK, preventing its simulation and
debugging on \pimsim, so we modified these three benchmarks to provide the same
functionality of the original implementation while being executable on top of
\pimsim. As discussed in \sect{sect:sim_dev},
the CPU$\leftrightarrow$DPU transfer (used for inter-DPU communication) is
modeled as a  fixed bandwidth model, so the frequency of inter-DPU
communications can affect the accuracy of \pimsim's simulated execution time.
To separately analyze the fidelity of \pimsim's DPU architecture model and the
effect CPU$\leftrightarrow$DPU communication model has on system-level
simulations, we separately report the validation results of \pimsim when
running the PrIM benchmark suite (1) with just a single DPU executing without
any inter-DPU communication and (2) with multiple DPUs with inter-DPU
communication.  For the single DPU validation, we used $710$ data points whose
execution times are within the range of $500$ ms, showing $98.4\%$ correlation
against \upmempim with a mean absolute error (MAE) of $12.0\%$.  Under the
multi-DPU validation, \pimsim shows $83.6\%$ correlation with MAE of $26.9\%$
under $387$ data points, with relatively larger absolute errors observed when
the inter-DPU communication time is more pronounced.

\subsection{Simulation Rate}
\label{sect:sim_perf_rate}

Developing a detailed execution-driven simulator often comes at the expense of
increased simulation time. While \pimsim is not multi-threaded, it achieves an
average simulation rate of $3$ KIPS (kilo-instructions-per-second), which is on
par with other popular execution driven simulators like
GPGPU-Sim~\cite{gpgpu_sim}. Because of UPMEM's current programming model and
how its communication/synchronization primitives work
(\sect{sect:background_pl}), DPUs mostly operate independently as a standalone
processor, so we expect parallelizing \pimsim with multi-threading will lead to
significant simulation rate improvements. We leave the support of
multi-threaded \pimsim implementation as future work.

\section{Demystifying \upmempim with \pimsim}
\label{sect:workload_characterization}

This section utilize \pimsim and the PrIM benchmark suite~\cite{prim_repo} to
demystify the internal runtime characteristics of \upmempim, 
showcasing the applicability of \pimsim for architectural exploration.  We
first focus on simulating PrIM under a single DPU setting in
\sect{sect:char_perf}, identifying its bottleneck in
\sect{sect:char_bottleneck}, and finally discussing multi-DPU execution with
strong scaling in \sect{sect:char_scaling}.  \tab{tab:prim_configurations}
summarizes the PrIM benchmarks and its input data sizes we explore in  this
paper. Due to space constraints, when sweeping the number of threads that
execute a given PrIM benchmark (collected over $1$/$2$/$4$/$8$/$16$/$24$
threads), we only show the results with $1$/$4$/$16$ threads for brevity.

\subsection{Analyzing Runtime Performance}
\label{sect:char_perf}

\fig{fig:ipc_and_bw_utilization} shows the compute and memory bandwidth
utilization as a function of the number of concurrent threads instantiated
($1$/$4$/$16$ threads).  With the exception of BS and SpMV, PrIM benchmarks
generally exhibit a compute-bound behavior, having a relatively higher compute
utilization than DRAM bandwidth utilization. PrIM targets data-intensive
workloads that are traditionally categorized as memory-bound under von-Neumann
CPU/GPU architectures. As such, the results in \fig{fig:ipc_and_bw_utilization}
highlight the unique value proposition of PIM vs. CPUs/GPUs, i.e., the
performance bottleneck is now shifted from the memory-bound regime to the
compute-bound territory.  We observe similar performance results over real
\upmempim systems (prior work in \cite{prim} reports similar observations),
demonstrating the fidelity of our \pimsim.

\subsection{Identifying Bottlenecks}
\label{sect:char_bottleneck}

While the workloads in PrIM generally exhibit a compute-bound behavior, the
results in \fig{fig:ipc_and_bw_utilization} imply that there are still some
performance left on the table.  Using \pimsim, we now root-cause the key
bottlenecks in \upmempim's microarchitecture that cause such performance loss.

{\bf Latency breakdown.} In \fig{fig:single_dpu_latency_breakdown}, we
breakdown DPU's execution time into two categories: (1) the time when the
thread scheduler has one or more threads to \emph{issue} into the pipeline
(black bar), and (2) when the scheduler is left idle with \emph{zero} threads
to issue (all non-black bars) because all the threads are either (2-a) waiting
for a memory operation to be serviced, (2-b) stalled due to the UPMEM's
revolver pipeline scheduling constraint, or (2-c) stalled due to the structural
hazard at the odd/even register file (see \sect{sect:background_hw_arch} for
revolver pipeline $\&$ RF hazard). As the number of threads increases, the DPU
scheduler is provided with more thread-level parallelism to  populate its
14-stage pipeline, leading to larger fraction of the runtime executing
instructions. Nonetheless, several workloads still suffer from non-negligible
portion of its execution time with idle cycles due to memory-side bottlenecks
(BS, SpMV), the structural hazards caused by the revolver pipeline and odd/even
RF (GEMV, HST-S, MLP, RED, TRNS, TS), or both (BFS, NW, SCAN-RSS, SCAN-SSA,
SEL, UNI). While pipeline stalls due to memory operations are a fundamental one
that cannot  be resolved easily through processor-side optimizations, idle
cycles originating from the revolver pipeline scheduling constraint or odd/even
RF hazard is an artifact of  \upmempim's specific processor microarchitecture. 

\begin{figure}[t!] \centering
  \includegraphics[width=0.485\textwidth]{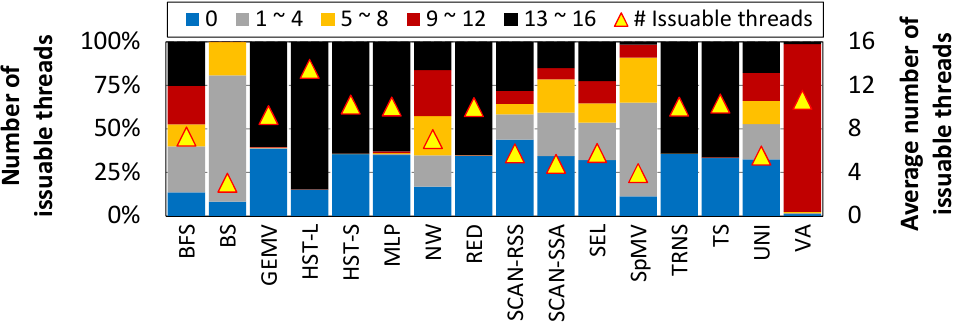}
\caption{Number of issuable threads by DPU scheduler each cycle, binned per each category (left axis) and average number of issuable threads (right axis) when executing with $16$ threads.}
\vspace{-1em}
\label{fig:number_of_active_threads_in_average}
\end{figure}

{\bf Thread-level parallelism (in space and time).} To analyze \upmempim's
performance bottleneck from a different dimension, we measure the amount of
thread-level parallelism (TLP) available to the DPU scheduler in space
(\fig{fig:number_of_active_threads_in_average}) and in time
(\fig{fig:number_of_active_tasklets_in_time_series}). In
\fig{fig:number_of_active_threads_in_average}, we categorize the number of
issuable threads available to the DPU scheduler to issue into the pipeline by
categorizing which bin it falls under.  As depicted, workloads suffering from
sub-optimal performance generally exhibit a higher fraction of its runtime with
less TLP (i.e., `$0$' issuable threads in the left axis of
\fig{fig:number_of_active_threads_in_average}), rendering the DPU to lose
compute throughput (\fig{fig:ipc_and_bw_utilization}). While insightful, the
analysis in \fig{fig:number_of_active_threads_in_average} cannot capture the
temporal variation in TLP or any phase behavior at runtime, which can add
another level of insights for architectural exploration. \pimsim enables the
analysis of how TLP fluctuates as execution progresses, as shown in
\fig{fig:number_of_active_tasklets_in_time_series}. Although some workloads
consistently exhibit low (BS) or high (GEMV) TLP, others exhibit a mix of
high-and-low TLP behavior (SCAN-SSA), providing valuable insights to understand
the runtime dynamics of a workload.

\begin{figure}[t!] \centering
\subfloat[BS]{\includegraphics[width=0.49\textwidth]{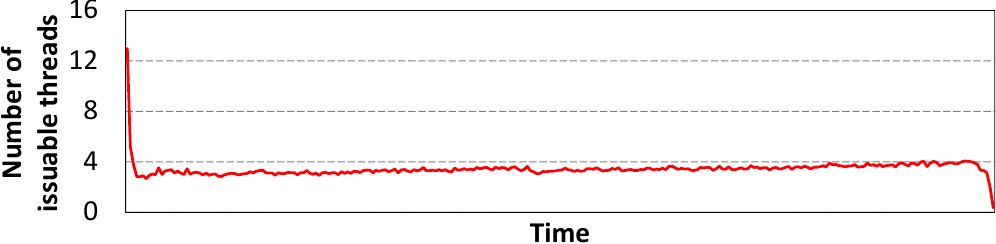}\label{fig:number_of_active_tasklets_in_time_series:bs}}\\
\subfloat[GEMV]{\includegraphics[width=0.49\textwidth]{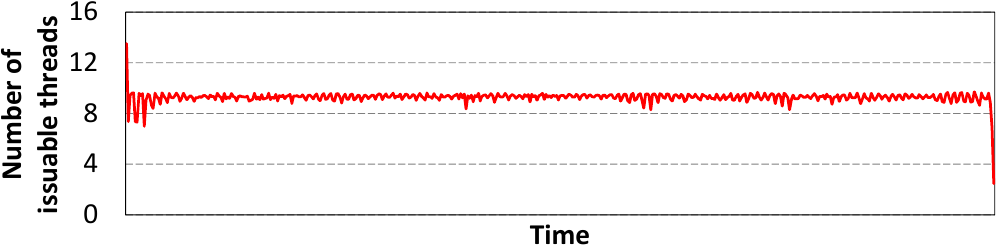}\label{fig:number_of_active_tasklets_in_time_series:gemv}}\\
\subfloat[SCAN-SSA]{\includegraphics[width=0.49\textwidth]
{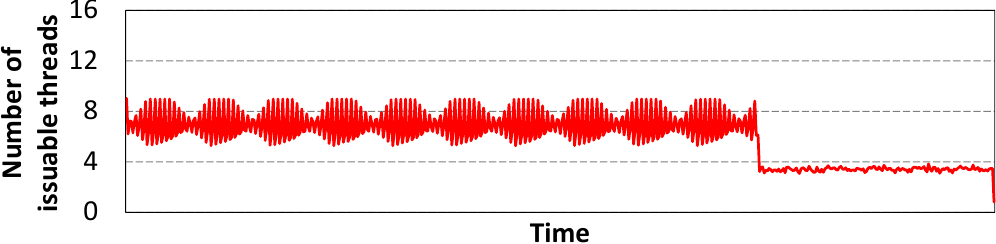}\label{fig:number_of_active_tasklets_in_time_series:scan-ssa}}
\caption{Changes in the number of issueable threads (y-axis) in time (x-axis) during the course of (a) BS, (b) GEMV, and (c) SCAN-SSA's execution. For clear visualization, the y-axis shows the number of issuable threads averaged over 10,000 consecutive cycles (i.e., cycles with \emph{zero} issuable threads are not shown clearly as they are smoothed out while averaging).}
\vspace{-1em}
\label{fig:number_of_active_tasklets_in_time_series}
\end{figure}

\begin{figure*}[t!] \centering
  \includegraphics[width=0.99\textwidth]{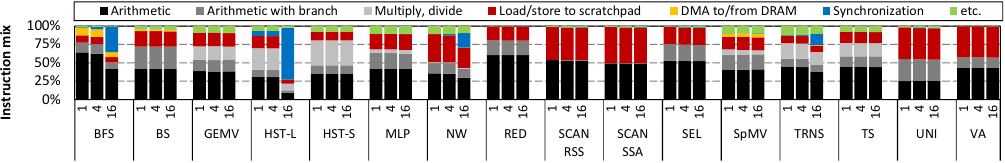}
\caption{Instruction mix when executing with a single DPU.}
\label{fig:instruction_mix}
\vspace{-0.5em}
\end{figure*}

\begin{figure*}[t!] \centering
  \includegraphics[width=0.99\textwidth]{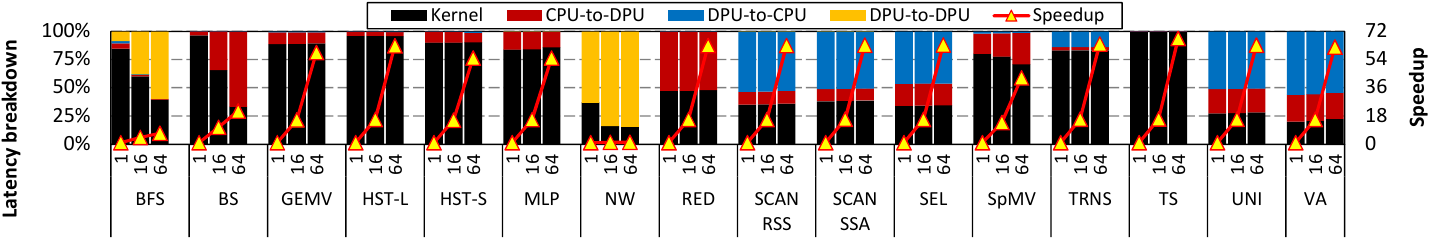}
\caption{Multi-DPU's latency breakdown (left axis) and achieved speedups (right axis) when strong-scaling PrIM across $1$/$16$/$64$ DPUs. All non-black bars represent communication latency. The DPU-to-DPU bar (yellow) in $1$ DPU shows latency incurred in copying input/output data in/out of the DPU across kernel execution boundaries.} 
\vspace{-1em}
\label{fig:multi_dpus_latency_breakdown}
\end{figure*}

{\bf Instruction mix. } Finally, \fig{fig:instruction_mix} shows the
instruction mix of PrIM when executed with a 1/4/16 threaded single DPU.
\pimsim uncovers a couple of interesting observations as follows. First, with
the exception of BFS, the number of load/store instructions to the scratchpad
memory (red) generally outweighs the number of DMA instructions to the per-bank
DRAM (yellow). 
This is an artifact of the scratchpad-centric programming model of \upmempim,
i.e., the register data operands can only be loaded from the scratchpad and the
programmer must manually initiate DRAM$\rightarrow$scratchpad copies to move
the working set closer to the processor. Consequently, to make sure the
scratchpad accesses do not cause a performance bottleneck, the DPU
microarchitecture is  designed to guarantee a short, single cycle latency in
handling load/store instructions.  Second,  although the compute utilization of
certain workloads like HST-L and TRNS seemingly look decent
(\fig{fig:ipc_and_bw_utilization}), a significant portion of its runtime is
effectively wasted as it is busy waiting to acquire locks (e.g.,
\texttt{mutex\_lock}). This is illustrated by the large fraction of the
instructions executed in HST-L and TRNS dedicated to synchronization
instructions (e.g., \texttt{acquire}, \texttt{release} in UPMEM ISA). Future
UPMEM ISA extensions that enable busy waiting threads to transition into sleep
mode and only resume execution when they are woken up can potentially reduce
such inefficiency.

\subsection{Strong Scaling with Multi-DPUs}
\label{sect:char_scaling}

\fig{fig:multi_dpus_latency_breakdown} shows the latency breakdown when each
PrIM benchmark is parallelized across $1$, $16$, and $64$ DPUs using
strong-scaling, i.e., benchmark's working set remains identical, so perfect
strong-scaling would reduce latency proportional to the number of DPUs.  In
general, the majority of PrIM's performance scales well when parallelized
across multiple DPUs because the communication size per DPU gets proportionally
reduced as a function of the DPUs concurrently executing. BFS, BS, and NW,
however, exhibit sub-linear scaling because the communication size gets larger
as the number of DPUs is increased. It is also worth noting that for some
benchmarks like SCAN-RSS, SCAN-SSA, SEL, UNI, and VA, transferring input
(CPU$\rightarrow$DPU) and output (DPU$\rightarrow$CPU) data dominates the total
execution time. For these benchmarks, the latencies to copy the input/output
data over CPU$\leftrightarrow$DPU channel are not being effectively hidden by
overlapping it with  DPU's kernel execution time.  Future versions of UPMEM SDK
which provide programming semantics that facilitate flexible kernel
partitioning and task scheduling (e.g., CUDA stream, CUDA dynamic
parallelism~\cite{cuda}) will likely enable further performance improvements. 

\section{Pathfinding Future PIM Architectures}
\label{sect:case_studies}

In this section, we uncover some key limitations of the current UPMEM-PIM
system through a series of case studies and demonstrate how \pimsim can be
utilized to explore architectural support required for \emph{future} PIM
architectures to become more performant, robust, and secure.

\subsection{Case Study $\#$1: UPMEM-PIM with SIMT Processing}
\label{sect:case_simt}

The baseline UPMEM-PIM employs a scalar processor leveraging
thread-level parallelism to maximize performance. Recent domain-specific
PIMs~\cite{hbm_pim_isca,aim}, on the other hand, leverage data-level
parallelism by employing vector processing to boost their performance for key
machine learning primitives (e.g., matrix multiplication). We observe that
UPMEM's SPMD programming model renders its hardware architecture to similarly
reap out performance benefits of data-parallel execution by employing a SIMT
(single-instruction-multiple-thread) vector processor~\cite{cuda}. In this
subsection, we augment the baseline UPMEM-PIM as follows to analyze the
performance benefits of employing SIMT vector processing. First, the processor
pipeline is augmented with a vector register file which an $N$-way vector unit
reads (write) vector operands from (to). Similar to the notion of ``warps'' in
CUDA, we group $N$ consecutive tasklets as the (grouped) thread scheduling
granularity to the $N$-way vector unit which executes $N$ scalar instructions
in lockstep for vector processing.  Similar to SIMT GPUs, a \emph{memory
address coalescing} operation ~\cite{cuda} is applied among the grouped $N$
scalar load/store instructions which helps maximize memory bandwidth
utilization by minimizing the effect of SIMT memory divergence
~\cite{chatterjee:mem_divg,meng:dws}. SIMT control
divergence~\cite{fung:dwf,fung:tbc,rhu:dpe,rhu:capri,rhu:slp,eltantawy:mpe,meng:dws}
is handled dynamically at runtime using each individual thread's program
counter values to only execute scalar threads executing the same instruction
over the vector lanes, masking out inactive threads from execution as
appropriate, similar to how recent NVIDIA GPUs (post Volta GPU) handle SIMT
control divergence~\cite{volta_independent_thread_sched}.

\begin{figure}[t!] \centering
\subfloat[]{\includegraphics[width=0.485\textwidth]{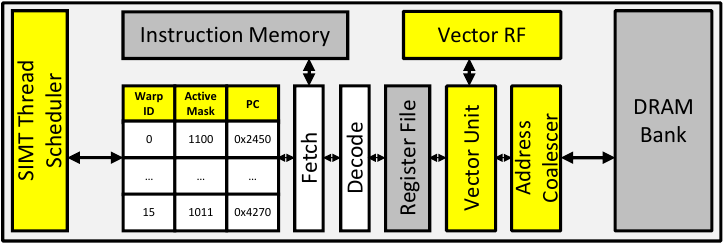}\label{fig:simd_architecture}}
\vspace{0em}
\subfloat[]{\includegraphics[width=0.49\textwidth]{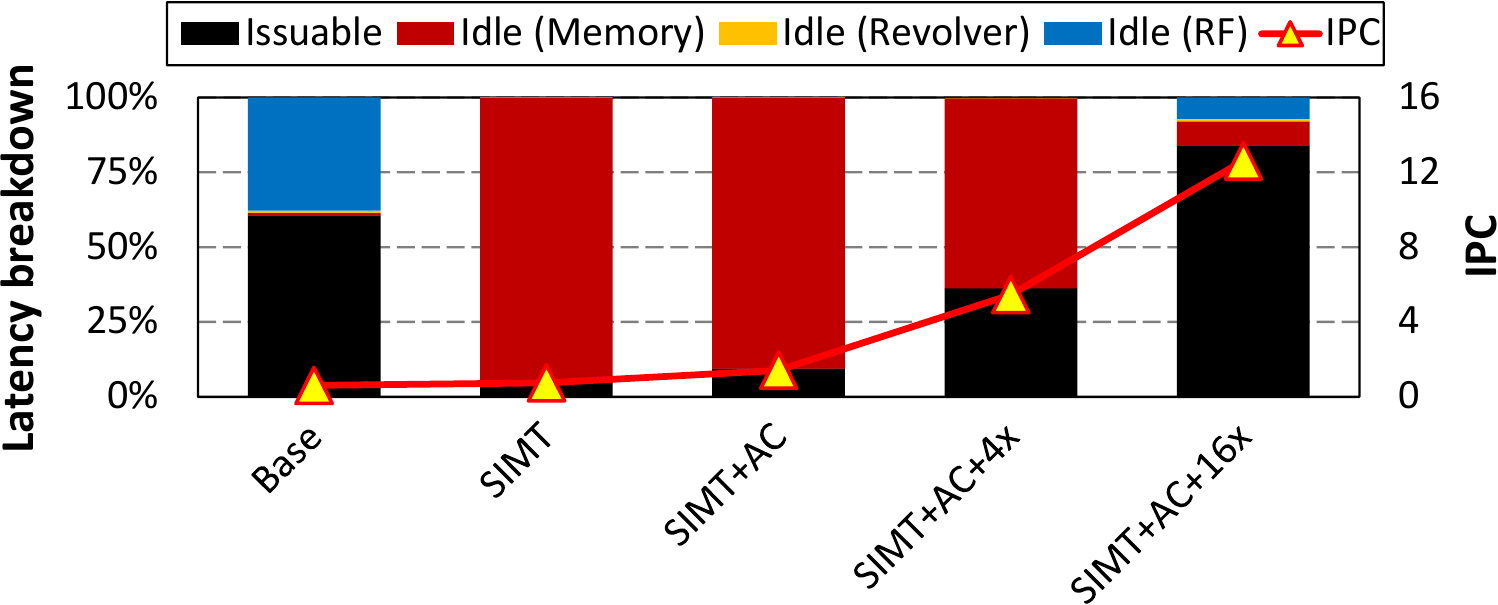}\label{fig:revision_simt_ipc_new}}
\caption{(a) SIMT based DPU architecture modeled using \pimsim, (b) performance (right axis) achieved for GEMV. The max IPC of \texttt{Base} and all \texttt{SIMT} designs are $1$ and $16$, respectively.}
\vspace{-1em}
\label{fig:revision_simt}
\end{figure}

\begin{figure*}[t!] \centering
\includegraphics[width=0.995\textwidth]{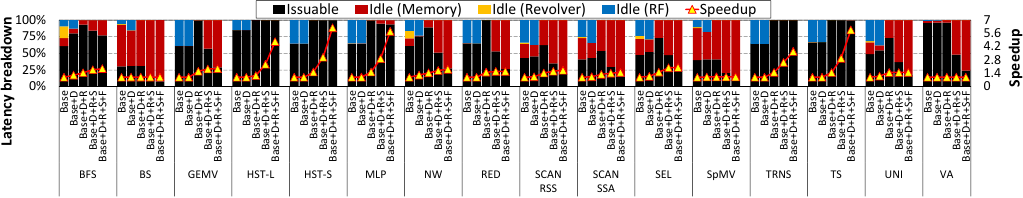} \caption{Ablation study to explore \upmempim's possible
  performance improvements when baseline DPU with $16$ threads is additively enhanced with data forwarding
  logic (D), unified RF with $2\times$ read bandwidth to remove hazards at RF (R), 2-way superscalar in-order pipeline (S), and
  doubling the operating frequency to $700$ MHz (F).} 
	\vspace{-1em} 
	\label{fig:case_study_ilp}
\end{figure*}

\fig{fig:revision_simt}  shows the performance achieved for GEMV, a key
primitive in machine learning which recent domain-specific PIMs are optimized
for. The figure first shows baseline UPMEM-PIM (\texttt{Base}), which is
additively augmented with 1) $16$-way SIMT vector unit \emph{without} memory
address coalescing (\texttt{SIMT}) and 2) \emph{with} address coalescing
(\texttt{SIMT+AC}). Both of these SIMT design points have MRAM read bandwidth
identical to \texttt{Base}.  Finally, another design point that scales up MRAM
read bandwidth by increasing DRAM operating frequency by $4\times$/$16\times$
(\texttt{SIMT+AC+4x/16x}) is explored to evaluate the upperbound performance
with SIMT.  As depicted, augmenting UPMEM-PIM with a $16$-way vector unit
(\texttt{SIMT}) provides a mere $2.6\times$ speedup as performance is heavily
bottlenecked by MRAM read bandwidth.  Adding the memory coalescer
(\texttt{SIMT+AC}) helps better utilize memory bandwidth and provides an
additional $1.9\times$ speedup vs. \texttt{SIMT} ($4.6\times$ vs.
\texttt{Base}).  Even with memory address coalescing (\texttt{AC}), however,
the memory-boundedness of SIMT execution persists with \texttt{SIMT+AC},
leaving performance left on the table, one which is only alleviated by the more
aggressive design which scales up MRAM bandwidth further with
\texttt{SIMT+AC+4x/16x}.

 \emph{ Key takeaways: UPMEM-PIM's SPMD programming model makes its hardware
architecture amenable to data-parallel processing via SIMT vector execution.
UPMEM-PIM's baseline memory system, however, is not sufficiently provisioned to
sustain the higher DRAM read bandwidth requirements of vector execution,
resulting in limited speedup with a naively implemented SIMT PIM design.
Properly optimizing the PIM memory system to maximize bandwidth utilization
(e.g., memory address coalescing, higher DRAM read bandwidth) will thus be
crucial for future SIMT vector based PIM designs to fully unlock the potential
of SIMT.  }

\begin{figure*}[t!] \centering
\includegraphics[width=0.99\textwidth]{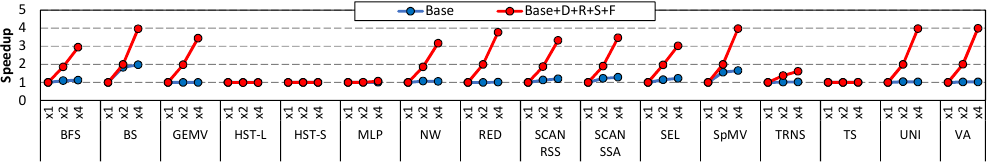} 
\caption{
Speedup achieved when scaling up the MRAM-to-WRAM bandwidth by four times ($\times1$ to $\times4$). The evaluated design points are 1) baseline UPMEM-as-is (Base) and 2) UPMEM with all the ILP optimizations (Base+(D+R+S+F)) discussed in \fig{fig:case_study_ilp}.
} 
\vspace{-1em}
\label{fig:revision_mram_scaling}
\end{figure*}

\subsection{Case Study $\#$2: ILP-enhanced PIM Architectures}
\label{sect:case_ilp}

Today's commercial PIM processors employ a simple, in-order pipeline without
any sophisticated microarchitectures to extract ILP for higher performance
(e.g., superscalar, super-pipelining, $\ldots$)~\cite{hbm_pim_isca,
hbm_pim_isscc, upmem_hotchips, newton}.  As discussed in
\fig{fig:single_dpu_latency_breakdown}, such a wimpy PIM processor design point
leaves significant performance left on the table, as conventionally
memory-bound workloads now fall under the compute-bound regime with PIM
(\sect{sect:char_perf}).  We believe such design decision was inevitable
because current generation of PIM processors are fabricated on a
density-optimized technology node (e.g., $\geq$20 nm DRAM technology for
HBM-PIM and \upmempim ~\cite{hbm_pim_isscc, prim}) posing several design
constraints that prevent advanced microarchitecture designs. That being said,
future PIM architectures with more flexibility in area, power, and thermal
budget can certainly consider relatively more aggressive, performance-oriented
design points with higher operating frequency and ILP-enhancing
microarchitectures.

In this case study, we use \pimsim to see how much performance can be unlocked
in PrIM's ``compute-bound'' workload by augmenting \upmempim's DPU with ILP
enhancing optimizations.  \fig{fig:case_study_ilp} summarizes our ablation
study on how much the baseline \upmempim's performance (denoted ``Base'') can
be improved by adding the following features in an additive manner: (D)
addressing the scheduling constraint enforced with baseline revolver pipeline
by enabling data forwarding across instructions without data dependencies
within the same thread to execute, (R) merging the odd/even RF into a single
one but doubling the read bandwidth to eliminate baseline RF's structural
hazard, (S) 2-way superscalar in-order pipeline to better leverage parallelism,
and finally (F) doubling DPU's operating frequency to $700$ MHz. As depicted,
the addition of these microarchitecture techniques substantially improve the
performance of PrIM's compute-bound workloads (avg $2.7\times$, max $6.2\times$
speedup) as it successfully addresses the performance bottlenecks discussed in
\fig{fig:single_dpu_latency_breakdown}.  Interestingly, with the addition of
(D+R+S) features to baseline UPMEM-PIM, several workloads become relatively
more memory-bound (i.e., larger fraction of Idle(Memory)) so the benefits of
higher operating frequency (F) are less pronounced for these workloads (e.g.,
GEMV, VA).  A fundamental reason why baseline UPMEM-PIM cannot fully reap out
the potential of these ILP optimizations is because of the large performance
gap between WRAM bandwidth ($2,800$ MB/sec) vs. MRAM-to-WRAM bandwidth
($600$-$700$ MB/sec). More concretely, for those workloads exhibiting low data
locality, the performance becomes relatively MRAM access bound and renders any
optimizations that resolve the compute-boundness of a workload ineffective
(e.g., all data points exhibiting high fraction of Idle(Memory) in
\fig{fig:case_study_ilp}).  Note that the existing $600$-$700$ MB/sec of
MRAM-to-WRAM bandwidth is not a fundamental constraint because the maximum
memory bandwidth that can be reaped out at the MRAM (DRAM) ``bank'' level is
much higher (up to several GB/sec of bandwidth), i.e., the limited $600$-$700$
MB/sec of MRAM bandwidth is simply a design point pursued by UPMEM-PIM
architects for this particular PIM design.  Using \pimsim, we conduct a
sensitivity study that scales \emph{up} the MRAM-to-WRAM read bandwidth and
analyze its performance implication for memory-bound workloads. As shown in
\fig{fig:revision_mram_scaling}, the speedup is more pronounced with the
ILP-enhanced UPMEM-PIM designs (red lines) because they exhibit more
memory-boundedness as shown in \fig{fig:case_study_ilp}.  Contrarily, benefit
of MRAM bandwidth scaling is minimal for workloads still exhibiting
compute-boundedness even under ILP optimizations (HST-L, HST-S, MLP, TRNS, TS).
Same principle holds for the baseline UPMEM as-is (blue lines) where the only
noticeable speedup with MRAM scaling is observed only for BS and SpMV which are
already heavily memory-bound even without ILP optimizations, the other
remaining compute-bound workloads achieving little speedup.

It is worth pointing out that, while the two case studies discussed so
far have quantified the performance merits of both SIMT and superscalar
execution in a PIM architecture, the available power and area budget can limit
how aggressively SIMT or superscalar can be employed within PIM. Standalone PIM
systems like SK Hynix's AiM~\cite{aim,aim_hotchips2023}, which are integrated
as co-processors on top of a PCIe bus, have much larger power and area budget
than a DIMM-based PIM solutions like UPMEM-PIM.  Therefore, these standalone,
domain-specific PIM solutions which have more design flexibility will more
likely be prime candidates to embrace SIMT or superscalar based PIM designs
that require higher design overheads.  

\emph{ Key takeaways: Many data-intensive workloads
	exhibit a compute-bound behavior when executed with PIM. Enhancing
PIM's computational throughput will therefore become much more important in
future PIM designs. Using \pimsim, we demonstrate the efficacy of various
ILP-enhancing microarchitectural techniques for future PIM architectures,
improving the performance of several compute-bound PIM workloads.  }

\subsection{Case Study $\#$3: Multi-tenant Execution in PIM}
\label{sect:case_multitenancy}

Multi-tenancy is one of the most important features for processors to provide
for cloud deployment as it  helps better saturate the processor's compute and
memory resources, reducing its total cost of ownership. As such, current
CPUs/GPUs come with a variety of hardware/software features that support
multi-tenancy with performance isolation and security guarantees (e.g., CPU
cache partitioning~\cite{arm_dynamlq, intel_mba}, NVIDIA's multi-instance
GPU~\cite{nvidia_mig}, etc.~\cite{linux_cgroup, linux_numactl}).  Given
\upmempim's abundant compute and memory throughput (e.g., an aggregate compute
and memory throughput of 0.896 TOPS and 2.5 TB/sec of memory bandwidth in a
$40$ ranked \upmempim system), having a proper multi-tenancy support will be
vital for future PIM architectures, especially when seeking for industrial
adoption by cloud vendors.

Unfortunately, our case study reveals that current commercial PIM chips
(whether it be \upmempim or domain-specific PIMs~\cite{hbm_pim_isca,
hbm_pim_isscc,newton,aim}) are not able to meet the requirements of
multi-tenancy, not just from a hardware/software  perspective, but also from
its programming model's perspective. Due to space limitations, let us focus our
attention on two important conditions to be met for multi-tenancy. First,
co-located workloads should securely execute without interfering with each
other (i.e., ``security'' guarantees). Second, co-located workloads must not be
aware of the fact that they are concurrently executing (i.e., ``transparency''
to co-located applications). We discuss each of these challenges below. 

\sethlcolor{cyan}

{\bf Security.} One of the fundamental architectural supports that is needed
for secure execution is isolating the memory address space of co-located
applications using MMU's \emph{address translation} capability. Practically all
commercial PIM systems~\cite{upmem_hotchips, hbm_pim_isca, hbm_pim_isscc,
newton} are designed \emph{without} an MMU, a point we emphasized in
\sect{sect:system_sw} with \upmempim's \emph{physical addressing} based memory
model.  Note that the decision regarding which granularity multi-tenancy should
be employed (e.g., coarse-grained per-DPU vs. fine-grained intra-DPU
multi-tenancy) presents interesting tradeoffs in terms of DPU resource
contention, virtualization overhead, etc. Such design decision, however, does
not obviate the need for the MMU to isolate different tenants by translating
virtual addresses. Consider a design point where per-DPU multi-tenancy is
implemented, e.g., two different PIM programs (two tenants) execute over
non-overlapping groups of DPUs within the same DIMM. If one of the tenants is a
malicious attacker, the malicious host-side CPU program can freely access the
other victim tenant's DPU physical memory because current PIM architectures
employ (MMU-less) physical addressing when accessing their DRAM banks.
Therefore, co-locating multiple workloads with address space isolation is
fundamentally impossible in MMU-less PIM architectures.

Aside from such security benefits, PIM chips with an MMU can greatly enhance
programmer productivity by \emph{virtualizing} the memory address space, i.e.,
they can separate the physical allocation of data in PIM DRAM against its
logical allocation within the virtual address space. As discussed in
\sect{sect:background_pl}, copying data from CPU to \upmempim's DRAM (MRAM)
requires the programmer to painstakingly derive the physical location in DPU's
DRAM because UPMEM ISA is currently based on physical addressing. Having a
proper MMU support will enable more flexible allocation of data across the
physical address space and can also provide ``pointer-is-a-pointer'' semantics
to enhance
programmability~\cite{cuda_uvm,supporting_x86_64_address_translation_for_100s_of_gpu_lanes,
architectural_support_for_address_translation_on_gpus, neummu}.  

In this case study, we add a hardware MMU to \upmempim, using our \pimsim, and
study its performance implications. Our MMU model employs a single page-table
walker (page size of $4$ KB) backed with a single-level, $16$-entry
fully-associative TLB.  The page-tables are assumed to be stored in DPU's local
DRAM bank and the access latency to the TLB is assumed as a single DPU clock
cycle.  Aside from how a PIM core (the DPU) handles address translation
exceptions, the interaction between a DPU and its MMU largely follows that of a
conventional CPU and its MMU.  That is, in the event that a DPU accesses
memory, the MMU translates all DRAM (MRAM)'s virtual address to its physical
address by leveraging the TLB or the page-table. For memory requests that the
MMU is not able to handle, however, an assistance from the host CPU is
required. This can occur, for example, when a page fault occurs and an update
to the page-table is in need. Under such circumstances, the MMU writes the
fault information into a fault buffer, which can either be recognized by the
host CPU via a polling approach or an interrupt-based approach. Under a polling
approach, the host periodically polls each DPU's fault buffer to fulfill DPU's
service needs. If the interrupt-based approach is to be employed, the MMU can
raise an interrupt-like signal via DDR4's \texttt{ALERT\_N} standard protocol
to interrupt and notify the host~\cite{micron_ddr4}. The host can then
recognize the existence of a page fault within the DPU and handle it
appropriately through a fault handler, updating the DPU's page-table before
sending a resume signal.  We utilize such mechanism to translate all DRAM
(MRAM) accesses from its  virtual address to its physical address and measure
its performance regression.  Overall, PrIM experiences an average performance
loss of $0.8\%$ (max $14.1\%$) by adding address translations to DPU's memory
accesses. Such low performance overhead is mainly attributed to UPMEM's
scratchpad-centric memory model where data transfers across
DRAM$\leftrightarrow$scratchpad are orchestrated in coarse-grained chunks
(several KBs) over DMA instructions, exhibiting high spatial locality and thus
achieving high TLB hit rates. Furthermore, DPU cores are clocked at $350$ MHz
frequency, rendering their memory access latency to be in the range of several
tens of DPU clock cycles (unlike CPUs/GPUs which operate in the GHz range and
exhibit hundreds of CPU/GPU processor cycles of memory latency), experiencing
much less TLB miss penalty than CPUs/GPUs. Overall, our case study demonstrates
the practicality of adding a functional MMU architecture to future PIM
technologies.

{\bf Transparency.} We believe that multi-tenant execution under the
\emph{current} UPMEM programming model is not practical because of its
scratchpad-centric programming model. Consider a scenario where we seek to
co-locate two PrIM benchmarks, a memory-bound BS and a compute-bound TS, which
exhibit complementary resource requirements (as quantified in
\fig{fig:ipc_and_bw_utilization}) and are perhaps the perfect candidates for
multi-tenant execution.  Unfortunately, the BS and TS each require using the
same scratchpad (WRAM)'s heap via a memory allocation API call
(\texttt{mem\_alloc()} in UPMEM SDK, e.g., line $35$-$37$ in
\fig{fig:upmem_programming}(b)), which leads to exceeding the total size of
scratchpad (WRAM)'s heap size. Consequently, co-locating both of these
workloads requires a non-trivial amount of changes to \emph{both} co-located
PrIM programs, arguably an unacceptable requirement to enforce on end-user
applications. More crucially, it directly violates the \emph{transparency}
requirement we previously discussed, rendering a scratchpad-centric PIM
programming model ill-suited for multi-tenant execution.

Consequently, our analysis reveals that future PIM should also employ
\emph{on-demand caches}, rather than singlehandedly relying on scratchpads, to
reap data locality benefits.  PIM programs running on top of an on-demand cache
will be able to leverage data locality in an opportunistic manner while not
having to change the program itself. In the next subsection, our next case
study details the feasibility of supporting on-demand cache architectures for
future PIM designs. 

\emph{ Key takeaways: Supporting multi-tenancy in PIM requires security and
transparency guarantees for the co-located workloads. To enhance security in
PIM architectures, we augment \pimsim with an MMU to quantify the performance
overheads of address translations and observe an average $0.8\%$ (max $14.1\%$)
latency increase, demonstrating the practicality of an MMU-enabled PIM design.
Guaranteeing transparency to co-located PIM workloads under UPMEM's current,
scratchpad-centric programming model is a different story, however, as it
requires significant changes to the co-located programs, a non-option to begin
with for transparent multi-tenant execution. Having an on-demand cache
architecture supported in PIM can bridge this gap, opening the door for
multi-tenant PIM architectures.  }

\begin{figure}[t!] \centering
  \includegraphics[width=0.485\textwidth]{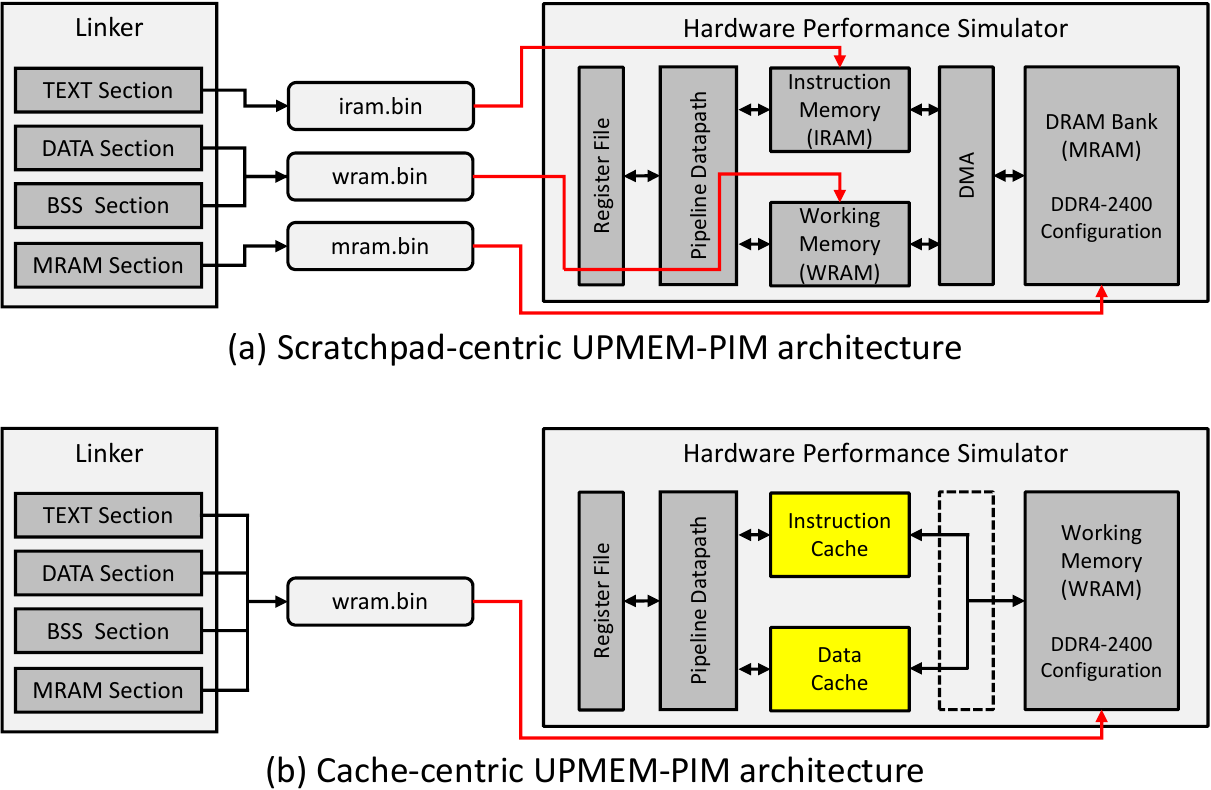}
\caption{Modeling a (a) scratchpad-centric and (b) cache-centric \upmempim architecture in \pimsim. }
\label{fig:case_study_caches}
\vspace{-1em}
\end{figure}

\begin{figure*}[t!] \centering
  \includegraphics[width=0.99\textwidth]{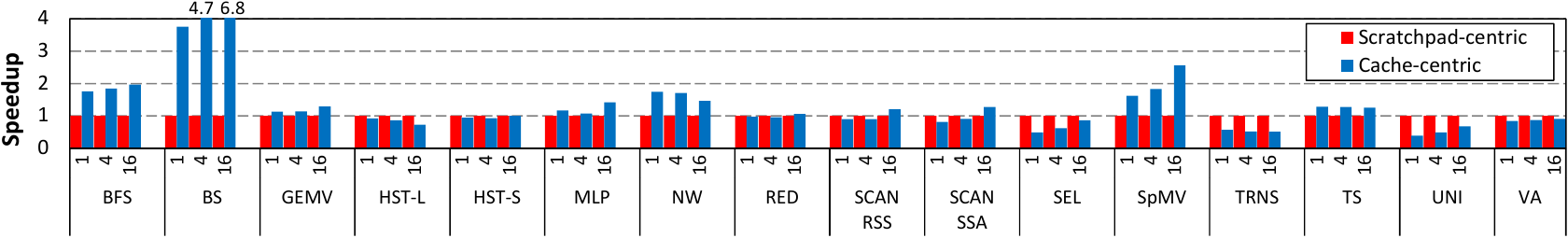}
\caption{Performance of scratchpad-centric vs. cache-centric \upmempim (normalized to scratchpad-centric design). The cache-centric UPMEM-PIM employs a 
	cache line size of $64$ bytes with load coalescing features enabled.}
\vspace{-0.7em}
\label{fig:execution_time_of_dpu_w_spm_vs_w_cache}
\end{figure*}

\subsection{Case Study $\#$4: On-demand Caches vs. Scratchpads}
\label{sect:case_caching}

As discussed in \sect{sect:background_pl}, UPMEM's scratchpad-centric
programming model requires the programmer to explicitly orchestrate the data
movements across two \emph{distinct} address spaces, the DRAM space and
scratchpad space (i.e., MRAM$\leftrightarrow$WRAM). This is because the DPU
threads can only load (store) data from (to) the scratchpad but cannot directly
address data mapped in the DRAM space. Using \pimsim's custom-designed linker,
this subsection conducts the cache vs. scratchpad case study based on the
following methodology.

\begin{enumerate}
    \item The open-source UPMEM compiler does not limit the data size the programmer can allocate and copy into 
      WRAM (scratchpad) space. Concretely, compiling an \upmempim program to an assembly-level code whose
      scratchpad allocation size exceeds the physical WRAM capacity in itself does not cause any compilation errors.
      During the linking process, however, if the WRAM data allocation size exceeds the \emph{physical} WRAM capacity,
      the UPMEM linker generates a linking error as the hardware \upmempim chip cannot execute the compiled codes
      properly (see \sect{sect:sim_dev} for discussion on UPMEM linker's key properties).
    
    \item Because \pimsim's linker is designed to  flexibly relocate and map a given address region to
      anywhere in the physical address space, we take the following measures to \emph{emulate} an
      alternative, cache-centric \upmempim (a) whose DPU threads can directly address data allocated in
      DRAM without having to move data to the scratchpad (i.e., there is no notion of scratchpad under this model),
      while (b) also leveraging data locality by storing recently accessed data within the cache.
 
    \item \pimsim emulates cache-centric \upmempim as follows. First, the input data is allocated directly in the WRAM
      (scratchpad) address space, unlike the baseline UPMEM model whose input data must be copied from MRAM (per-bank
      DRAM) to WRAM using DMA instructions. The WRAM-allocated input data, which is directly addressable by the DPU
      threads using load/store instructions (as compiled by the original UPMEM compiler), is then relocated by \pimsim's
      linker to be mapped into a physical address region which is backed by our cycle-level hardware performance
      simulator, modeling it as a DDR4-2400~\cite{ddr4_2400} compatible DRAM device (\fig{fig:case_study_caches}(b)).
      By adding a cycle-level cache simulator in between the DPU processor and the (DRAM-emulated) WRAM address space,
      the data referenced by the load/store instructions will be stored on-demand to this cache simulator, allowing us
      to explore the cache vs. scratchpad design space.
     
\end{enumerate}

\fig{fig:execution_time_of_dpu_w_spm_vs_w_cache} compares the performance of
scratchpad vs. cache in \upmempim for PrIM.  The cache-centric \upmempim
employs an instruction cache and a data cache, each configured as an $8$-way
set-associative cache with LRU replacement policy and $24$ KB and $64$ KB
capacity, respectively, identical to the instruction memory (IRAM) and
scratchpad (WRAM) space provisioned under the baseline \upmempim. For certain
workloads, scratchpad performs better than caches (e.g., UNI) while the
opposite also holds true for others (e.g., BS). To better understand the
reasons behind such results, \fig{fig:cache_vs_spm_bytes_read} shows the number
of bytes read from DRAM during the course of BS and UNI's execution.  In
general, we can observe that the execution time is highly correlated with the
number of bytes read. For example, under the memory-bound BS, the scratchpad
based execution with 16 threads incurs $5.1\times$ higher memory read traffic
than using caches. For BS, it is challenging to statically estimate the right
amount of data to upload into the scratchpad, which results in a severe
\emph{overfetching} of useless data and causing a performance bottleneck to
this memory-bound workload. Under such scenario, a purely on-demand caching
strategy performs much more favorably in terms of fetching (relatively) the
right amount of data within the cache and maximizes data locality.  In
contrast, workloads like UNI performs much better with scratchpads where
carefully orchestrating data movements perform better than the opportunistic
cache design. Determining which design point is more favorable for PIM
architectures is not the purpose of this case study. Rather we seek to
demonstrate the practical benefits and feasibility of a cache-centric PIM
architecture, motivating future work in this research space.

\emph{ Key takeaways: Similar to conventional CPUs/GPUs, an on-demand cache
design can do a better job in leveraging locality for PIM when its memory
access pattern cannot be optimally determined at compile time, a scenario where
scratchpad based design points can perform poorly.  }

\begin{figure}[t!] \centering
  \includegraphics[width=0.495\textwidth]{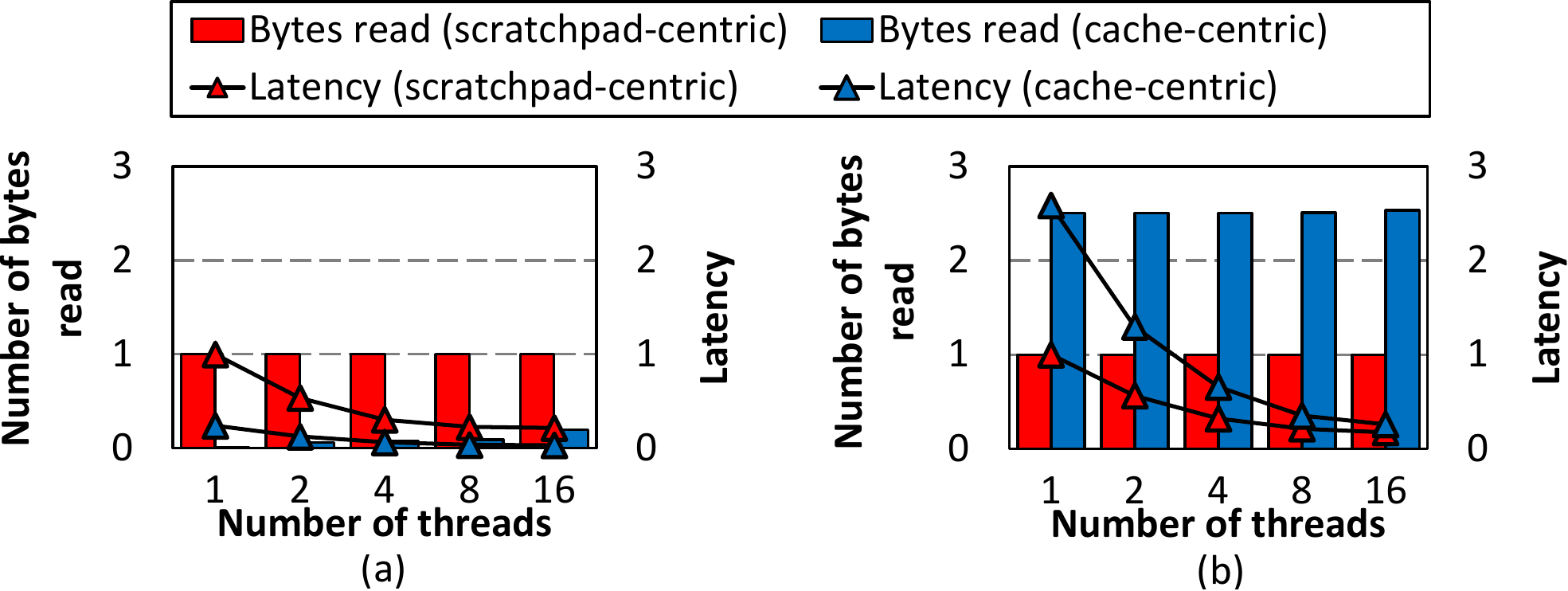} 
	\vspace{-1.5em} 
	\caption{
  Bytes read from DRAM (left axis, normalized) and end-to-end execution time (right axis, normalized) for (a) BS and
  (b) UNI.} 
	\vspace{-1em}
	\label{fig:cache_vs_spm_bytes_read}
\end{figure}

\subsection{Other Promising Research Directions}
\label{sect:case_misc}

Aside from the case studies we discussed previously, we believe that PIM with
better inter-DPU synchronization primitives, high-performance inter-DPU
communications, robust system software support for better programmability, and
a unified virtual memory across all DPUs are critical components that require
attention from PIM architects. We plan on exploring these studies as future
work.

\section{Simulator Limitations and Future Work}
\label{sect:future}

\subsection{Power and Area Modeling for PIM}

Similar to the early efforts on modeling cycle-level performance of
programmable GPUs~\cite{gpgpu_sim}, the current version of \pimsim primarily
focuses on simulating the performance aspects of UPMEM-PIM.  There exists a
rich set of prior work focusing on estimating the power and area of
CPUs~\cite{wattch,mcpat} and
GPUs~\cite{gpuwattch,accelwattch,isca2010:An_Integrated_GPU_Power_and_Performance_Model}
and integrating them with cycle-level CPU/GPU performance model simulators.  An
important future direction of \pimsim is to develop a power and area modeling
framework targeting PIM and integrate them with our UPMEM-PIM performance
model.  We leave it as  future work as it deserves a detailed exploration on
its own.

\subsection{Improving the Fidelity of Inter-DPU Communication}

As discussed in \sect{sect:sim_validation}, using a simple bandwidth model for
CPU$\leftrightarrow$DPU communications renders \pimsim to exhibit relatively
lower correlation with real UPMEM-PIM system when the inter-DPU communication
time is more pronounced. A real UPMEM-PIM system implements such communication
operator by having the host CPU execute AVX instructions, so improving the
fidelity of \pimsim's inter-DPU communication requires our simulation framework
to be tightly integrated with a detailed cycle-level CPU performance
model~\cite{gem5,zsim,multi2sim,sniper,graphite,ptlsim}.  Extending \pimsim to
be integrated with high fidelity CPU simulators is left as  future work.

\section{Related Work}
\label{sect:related_work}

The initial concept of PIM dates back to the
1970s~\cite{a_logic_in_memory_computer} with numerous follow-on
works~\cite{computational_ram,iram,active_pages,diva}. With the proliferation
of today's domain-specific architectures, there exists a variety of PIM or
near-memory processing studies~\cite{graph_p, graph_h, graph_q, scalagraph,
tesseract, hbm_pim_isca, axdimm, newton, tetris, neurocube, tensor_dimm,
recnmp, centaur, trim_micro, trim_cal, gradpim, fafnir, impica, polynesia,
jafar, dracc, cmp_pim, isaac, dimmining, spacea, gearbox, natsa, drisa, mcdram,
mcdram_v2, medal,
accelerating_bandwidth_bound_deep_learning_inference_with_main_memory_accelerators,
an_fpga_based_rnn_t_inference_accelerator_with_pim_hbm, chameleon,
near_dram_acceleration_with_single_isa_heterogeneous_processing_in_standard_memory_modules,
nda,
application_transparent_near_memory_processing_architecture_with_memory_channel_network,netdimm}.
There are also several prior works on PIM exploring compiler
issues~\cite{to_pim_or_not, tom, cairo}, data
coherency~\cite{pim_enabled_instructions, lazypim, conda,
near_data_acceleration_with_concurrent_host_access},
synchronization~\cite{syncron}, QoS aware runtime and scheduling for
PIM~\cite{pimcloud}, among many others~\cite{ambit, d_range, quac_trng,
simdram, elp2im, rowclone, lisa}.  This paper focuses on characterizing the
first real-world general-purpose PIM via our \pimsim, pathfinding important
research directions for future PIMs.  Below we summarize other relevant works
on characterizing real-world PIM and PIM simulators. 

{\bf Analysis on real-world PIM devices.} There have been several recent works
that characterize commercial PIM technologies~\cite{prim, prim_repo,
hbm_pim_isca, benchmarking_memory_centric_computing_systems, sparsep,
accelerating_neural_network_inference_with_processing_in_dram,
a_case_study_of_processing_in_memory_in_off_the_shelf_systems,
an_experimental_evaluation_of_machine_learning_training_on_a_real_processing_in_memory_system,
variant_calling_parallization_on_processor_in_memory_architecture,
dna_mapping_using_processor_in_memory_architecture,
bulk_jpeg_decoding_on_in_memory_processors}. Gómez-Luna et al.~\cite{prim,
prim_repo} developed the PrIM benchmark suite and conducted a  workload
characterization on \upmempim. There are also several works exploring the
applicability of \upmempim for accelerating dense/sparse linear algebra,
databases, data analytics, graph processing, bioinformatics, image processing,
compression, simulation, encryption, and etc~\cite{prim, prim_repo,
benchmarking_memory_centric_computing_systems, sparsep,
a_case_study_of_processing_in_memory_in_off_the_shelf_systems,
variant_calling_parallization_on_processor_in_memory_architecture,
dna_mapping_using_processor_in_memory_architecture,
bulk_jpeg_decoding_on_in_memory_processors}, with more recent work exploring
\upmempim's applicability for accelerating machine
learning~\cite{an_experimental_evaluation_of_machine_learning_training_on_a_real_processing_in_memory_system}.
Lee et al.~\cite{hbm_pim_isca} discusses the hardware/software architecture for
Samsung's HBM-PIM architecture. There is also a recent work by Liu et
al.~\cite{axdimm} which explores the applicability of Samsung's near-memory
processor AxDIMM for accelerating recommendation models.

{\bf Simulators for PIM.}   PIMSim~\cite{pimsim} supports a configurable PIM
logic modeling, memory organization, interconnection, and provides
co-simulation with other simulation frameworks.  Ramulator-PIM~\cite{ramulator,
ramulator_pim_repo, napel} integrates ZSim~\cite{zsim} and
Ramulator~\cite{ramulator} to simulate PIM-enabled memory.
MPU-Sim~\cite{mpusim} models a near-bank processing architecture which supports
NVIDIA CUDA's SIMT programming model~\cite{cuda}.  MultiPIM~\cite{multipim}
enables the simulation of PIM systems based on 3D stacked memory with features
to explore multi-stack interconnects with virtual memory support.  Compared to
these existing PIM simulators, the key novelties of \pimsim are as follows.
First, the frontend of our software compilation toolchain employs a
custom-designed linker targeting industry's first general-purpose PIM ISA,
which opens up a wide range of hardware/software architectural explorations.
Existing PIM simulators primarily rely on conventional software frontends
(e.g., x86 in ZSim+Ramulator), making it challenging to change the way the
program and data binaries are mapped over the memory address space, a feature
critical in some of the case studies we conducted in \sect{sect:case_studies}.
Second, \pimsim's backend simulator models a real-world per-bank PIM
architecture, widely employed and commercialized in both
domain-specific~\cite{hbm_pim_isca,aim} and general-purpose PIM designs, unlike
popular PIM simulators like MultiPIM or
ZSim+Ramulator~\cite{multipim,ramulator, ramulator_pim_repo, napel} which
assume the PIM cores are placed in the logic layer of a 3D stacked memory
(e.g., HMC).  \tab{tab:revision_pim_vs_others} summarizes key differences
between \pimsim and other PIM simulators.

\begin{table}[t!] \centering
  \footnotesize
    \caption{Comparison of \pimsim vs. other PIM simulators.}
  \resizebox{\columnwidth}{!}{
  \begin{tabular}{|c|c|c|c|c|c|c|}
  \hline
   &
    \textbf{\begin{tabular}[c]{@{}c@{}}PIMSim\\ \cite{pimsim}\end{tabular}} &
    \textbf{\begin{tabular}[c]{@{}c@{}}Ramulator\\-PIM~\cite{ramulator_pim_repo}\end{tabular}} &
    \textbf{\begin{tabular}[c]{@{}c@{}}MultiPIM\\ \cite{multipim}\end{tabular}} &
    \textbf{\begin{tabular}[c]{@{}c@{}}MPU-Sim\\ \cite{mpusim}\end{tabular}} &
    \textbf{\pimsim} \\ \hline
  \textbf{ISA} &
    \begin{tabular}[c]{@{}c@{}}x86, ARM,\\ SPARC\end{tabular} &
    x86 &
    x86 &
    PTX &
    UPMEM \\ \hline
  \textbf{Implementation} &
    \begin{tabular}[c]{@{}c@{}}In-house\end{tabular} &
    \begin{tabular}[c]{@{}c@{}}Zsim\\ + Ramulator\end{tabular} &
    \begin{tabular}[c]{@{}c@{}}Zsim\\ + Ramulator\\ + BookSim\end{tabular} &
    \begin{tabular}[c]{@{}c@{}}In-house \end{tabular} &
    \begin{tabular}[c]{@{}c@{}}In-house \end{tabular}
    \\ \hline
  \textbf{\begin{tabular}[c]{@{}c@{}}Frontend \\(Trace vs. Execution)\end{tabular}} &
    \begin{tabular}[c]{@{}c@{}}Trace\end{tabular} &
    \begin{tabular}[c]{@{}c@{}}Trace\\ + Execution\end{tabular} &
    \begin{tabular}[c]{@{}c@{}}Trace\\ + Execution\end{tabular} &
    \begin{tabular}[c]{@{}c@{}}Execution \end{tabular} &
    \begin{tabular}[c]{@{}c@{}}Execution \end{tabular}
    \\ \hline
  \textbf{\begin{tabular}[c]{@{}c@{}}ISA \& Linker \\ Customization\end{tabular}} &
    \cellcolor[HTML]{FFCCC9}X &
    \cellcolor[HTML]{FFCCC9}X &
    \cellcolor[HTML]{FFCCC9}X &
    \cellcolor[HTML]{FFCCC9}X &
    \cellcolor[HTML]{D3F5C7}O \\ \hline
  \textbf{\begin{tabular}[c]{@{}c@{}}Validation Against\\Real  PIM Hardware\end{tabular}} &
    \cellcolor[HTML]{FFCCC9}X &
    \cellcolor[HTML]{FFCCC9}X &
    \cellcolor[HTML]{FFCCC9}X &
    \cellcolor[HTML]{FFCCC9}X &
    \cellcolor[HTML]{D3F5C7}O \\ \hline
  \textbf{\begin{tabular}[c]{@{}c@{}}Multi-threaded \\ Simulation\end{tabular}} &
    \cellcolor[HTML]{FFCCC9}X &
    \cellcolor[HTML]{D3F5C7}O &
    \cellcolor[HTML]{D3F5C7}O &
    \cellcolor[HTML]{FFCCC9}X &
    \cellcolor[HTML]{FFCCC9}X \\ \hline
    \textbf{\begin{tabular}[c]{@{}c@{}}Lines of Code \\ (LoC)\end{tabular}} &
    30 K &
    35 K &
    92 K &
    12 K &
    52 K \\ \hline
    \textbf{\begin{tabular}[c]{@{}c@{}}Simulation Rate \\ (KIPS)\end{tabular}} &
    N/A &
    N/A &
    N/A &
    N/A &
    3 \\ \hline
  \end{tabular}%
  }
\vspace{-1.2em}
\label{tab:revision_pim_vs_others}
  \end{table}

\section{Conclusion}
\label{sect:conclusion}

In this work, we present a novel simulation framework named \pimsim which
targets UPMEM's commercial general-purpose PIM architecture. Using \pimsim, we
present our detailed characterization on wide range of real PIM programs and
showcase \pimsim's applicability for computer architecture research.
Furthermore, we identify some important shortcomings of the current UPMEM-PIM
system through our case studies and propose some critical research areas that
require further investigation from computer architects (e.g., architectural
support for vector processing, ILP-enhancing microarchitectures, multi-tenancy,
and on-demand caching), which we believe will be vital for future PIM
architectures to evolve into first class computing citizens.

\section*{Acknowledgment}
This research is supported by
the National Research Foundation of Korea (NRF) grant funded by the Korea government(MSIT) (NRF-2021R1A2C2091753),
Institute of Information \& communications Technology Planning \& Evaluation (IITP) grant funded by the Korea government(MSIT) (No. 2022-0-01037, Development of High Performance Processing-in-Memory Technology based on DRAM),
and by Institute of Information \& communications Technology Planning \& Evaluation (IITP) grant funded by the Korea government(MSIT) 
(No.2019-0-00075, Artificial Intelligence Graduate School Program(KAIST)). We also appreciate the support from 
Samsung Electronics (Samsung-KAIST Center for Memory-Centric System Architecture) and 
Samsung Electronics Co., Ltd (IO201210-07974-01).

\bibliographystyle{ieeetr}
\bibliography{refs}

\end{document}